\documentclass[aps,twocolumn,prl,footinbib,longbibliography]{revtex4-1}
 \pdfoutput=1
\usepackage{amsmath}
\usepackage{amsfonts}
\usepackage[utf8]{inputenc}
\usepackage{graphicx}
\usepackage{xcolor}
\usepackage[colorlinks=true,linkcolor=blue,urlcolor=blue,citecolor=blue]{hyperref}
\usepackage{color}

\newcommand{\PathToFigures}{}

\begin{document}

\title{Disorder-driven quantum transition in relativistic semimetals:
functional renormalization via the porous medium equation
}

\author{Ivan Balog$^1$,  David Carpentier$^2$,  and  Andrei A. Fedorenko$^2$}
\affiliation{$^1$Institute of Physics, P.O.Box 304, Bijeni\v{c}ka cesta 46, HR-10001 Zagreb, Croatia  \\
$^2$\mbox{ Universit\'{e} de Lyon, ENS de Lyon, Universit\'{e}  Claude Bernard, CNRS, Laboratoire de Physique, F-69342 Lyon, France} }

\date{October 19, 2018}

\begin{abstract}
In the presence of randomness, a relativistic semimetal undergoes a quantum
transition towards a diffusive phase. A standard approach relates this transition to the
$U(N)$ Gross-Neveu model in the limit of $N \to 0$.
We show that the corresponding fixed point is infinitely unstable,
demonstrating the necessity to include fluctuations beyond the usual Gaussian approximation.
We develop a functional renormalization group method amenable to include these effects  and
show that the  disorder distribution  renormalizes following
the so-called porous medium equation. We find that
the transition is controlled by a nonanalytic fixed point drastically
different from that of the $U(N)$ Gross-Neveu model.
Our approach provides a unique  mechanism of spontaneous generation of a finite density of states
and also characterizes the scaling behavior of the broad distribution
of fluctuations close to the transition.
It can be applied to other problems
where nonanalytic effects may play a role, such as the Anderson localization transition.

\end{abstract}

\maketitle

\textit{Introduction. --}
The interplay between disorder and quantum fluctuations leads to unique phenomena, the
most remarkable being the Anderson localization.
After more than half a century of intensive efforts, it remains a topical subject of research with
applications to various domains of physics ranging from condensed matter to cold atoms and
light propagation~\cite{Abrahams:2010}.
Remarkably, a different type of disorder-driven  quantum phase transition was discovered recently
when considering waves with a quantum relativistic dispersion relation~\cite{Syzranov:2018}.
This transition happens between a pseudoballistic phase and a diffusive metal
as a function of the disorder strength (or the energy).
It is predicted to occur in particular in the recently discovered
three-dimensional (3D)  Weyl~\cite{Xu:2015a,Xu:2015b} and
Dirac semimetals~\cite{Liu:2014,Neupane:2014,Borisenko:2014}
in which, respectively, two and four electronic bands cross linearly at isolated points. However we expect
these phenomena to be relevant to other relativistic waves beyond condensed matter, such as
ultracold atoms~\cite{Dubcek:2015}.

In spite of numerous efforts, the understanding of this transition remains elusive.
In this Letter we show that the fluctuations of the randomness beyond the standard Gaussian approximation
invalidate previous field-theoretic descriptions of this transition.
A very similar mechanism occurs in the context of the Anderson transition:
there discrepancies between the results obtained using renormalization group (RG) and numerical simulations grow
with the number of loops~\cite{Slevin:1999}.
One may attribute them to the existence of infinitely many relevant
operators of the associated field theory~\cite{Evers:2008} which destabilize the fixed point (FP)
usually considered to describe the transition~\cite{Kravtsov:1989,Wegner:1990,Castilla:1993}.
We find that the same  problem appears at the new semimetal-diffusive metal transition.
We demonstrate how to overcome this obstacle by deriving and solving a functional renormalization group
(FRG) for the whole (non-Gaussian) disorder distribution.  To our knowledge this solution constitutes
the only example of an analytical description of a disorder-driven quantum phase transition controlled
by non-Gaussian disorder fluctuations.
Besides the present work  we are aware of only one other example, namely the classical
 2D \textit{XY} model with random phases, for which not only the flow
equation for the probability distribution but also its solution was obtained by a mapping to the so-called
Kolmogorov-Petrovskii-Piscounov equation~\cite{Carpentier:1998,Horovitz:2002}.
Hence we believe that our
work sheds new light on the description of critical non-Gaussian disorder fluctuations in quantum systems
beyond the disorder-driven semimetal-diffusive metal phase transition.

Here we focus on the transition between a pseudoballistic semimetal phase
with a vanishing  density of states (DOS) at the nodal point
and a diffusive metal phase with a finite DOS at zero
energy~\cite{Fradkin:1986,Roy:2016b,Sbierski:2014,Goswami:2011,Hosur:2012,Ominato:2014,Chen:2015,Altland:2015:2016}.
The field-theoretic description of this transition
using both replica and SUSY
approaches~\cite{Syzranov:2015b,Roy:2014,Louvet:2016,Louvet:2017}
implies  that in the absence of scattering between different nodal points the transition
is controlled   by a perturbative in  $\varepsilon=d-2$  FP
of the $d$-dimensional $U(N)$ Gross-Neveu (GN) model taken in the
unusual limit of a vanishing number of fermion flavors $N\to 0$.
As for the Anderson transition, numerical studies~\cite{Kobayashi:2014,Sbierski:2015,Liu:2016,Fu:2017,Sbierski:2017}
demonstrate quantitative discrepancies with the predictions of the
GN model~\cite{Syzranov:2015b,Roy:2014,Louvet:2016} which grow with the
order of approximation.
We show that the GN FP is infinitely unstable in the limit of $N\to 0$ implying that
the non-Gaussian fluctuations of the randomness are at the origin of the breakdown  of the GN description.
It is the purpose of the present Letter to resolve this problem by
deriving the flow equation for the whole
disorder distribution
and solving it through a mapping to the well-known
porous medium equation~(PME)~\cite{Vazquez:2007}.
This reveals that the phase transition is governed by
a nonanalytic FP  which  is crucially different from that of the GN model.

\textit{Model. --}
We start from the imaginary time action of relativistic
fermions moving in a $d$-dimensional space in the presence of an
external potential $V(r)$
\begin{equation} \label{eq:action-Weyl-3D-time}
  S= i \int d^d x d\tau\, \bar{\psi}(x,\tau)[\partial_\tau - i  \gamma_j \partial_j+V(x)]\psi({x,\tau}),
\end{equation}
where $\bar{\psi}$ and $\psi$ are independent Grassmann fields and $\tau$ is the imaginary time.
$\gamma_j$ are elements of a Clifford algebra   satisfying the anticommutation relations:
$\gamma_j \gamma_k + \gamma_k \gamma_j = 2 \delta_{jk} \mathbb{I}$ ($j,k=1,...,d$),
which reduce to the Pauli matrices $\gamma_j = \sigma_j$ in $d=3$.
The disorder potential is assumed to be uncorrelated in space, and thus,
its distribution can be described by a
local characteristic function $W(\Theta)$ defined as
$ \overline{\exp(- i \int d^d x V(x)\Theta(x) )} = \exp[ - \int d^d x  W(\Theta(x)) ]$.
Here the overbar stands for averaging over disorder configurations.
To perform averaging directly in the action~(\ref{eq:action-Weyl-3D-time})
we use the replica trick.
Since the fermions are noninteracting it is convenient to switch in the
action~(\ref{eq:action-Weyl-3D-time})
from the imaginary time to the Matsubara frequency and write down
the bare replicated action at
fixed energy $\omega$ as~\cite{Ludwig:1994}
\begin{equation} \label{eq:action-Weyl-replicated}
  S=  \int d^d x \sum\limits_{\alpha=1}^{N}  \bar{\psi}_\alpha (x)(
   \gamma_j \partial_j +  \omega )\psi_\alpha(x) + W(\Theta(x)),
\end{equation}
where $\Theta(x) = \sum_{\alpha=1}^{N} \bar{\psi}_\alpha (x) \psi_\alpha (x)$
is the local density of fermions.

\textit{Renormalization. -}
To derive the FRG flow equations we use the effective average action formalism developed by
Wetterich~\cite{Wetterich:1991} together with $\varepsilon=d-2$ expansion.
Introducing the IR cutoff in the form of mass $m$
we obtain the flow equation for the characteristic function~\cite{Supplementary}
\begin{eqnarray}
\label{eq:perturb-unrescaled}
-m \partial_m W(\Theta)&=&  2 m^{\varepsilon} \left( \Theta W'(\Theta)W''(\Theta)  -
\tilde{N} W'(\Theta)^2  \right ), \ \ \ \
\end{eqnarray}
where  $\tilde{N} = \frac{N}2 \mathrm{tr} \mathbb{I} $,
 $\Theta$ is the expectation value of $\Theta(x)$, and $m$ goes from $m_0$ to $0$.
A counterpart of Eq.~(\ref{eq:perturb-unrescaled}) derived in a fixed dimension $d$
can be found in~\cite{Supplementary}.
%%%%%%%%%%
\nocite{Berges:2002,Hofling:2002,Rosa:2001,Mouhanna:2010,Delamotte:2012,%
Jakovac:2013,Zinn-Justin:1986,LeDoussal:2004,Fedorenko:2006,Bec:2007,%
Aoki:2014,Evans:1998,Angenent:2001,Betelu:2000%,Horovitz:2002,Louvet:2016,Vazquez:2007,Gratton:1990,Aronson:2003,Wiese:2016
}
%%%%%%%%%%
The renormalized Green's function corresponding to action~(\ref{eq:action-Weyl-replicated}) is
$G_{\alpha \beta}(k) = \delta_{\alpha \beta}/[\gamma_j k_j -i\omega - iW'(0) ]$.
For physically relevant disorder distributions
the bare characteristic function $W(\Theta)$ is analytic and satisfies $W'(0)=0$.
Hence, the bare DOS given by
$\rho(\omega) = - 1/\pi {\mathrm{Im}} \int_k G_{\alpha \alpha}(k,\omega) $
vanishes at zero energy~\cite{Pixley:2017}.
However, as we will see later the renormalized characteristic function can develop a cusp
at the origin, and thus, generate  a nonvanishing DOS at zero energy.

To demonstrate how one can recover the FP of the $U(N)$ GN model
we rewrite the FRG equation~(\ref{eq:perturb-unrescaled}) in dimensionless
form by substituting $W(\Theta) = m^{2+\varepsilon} w (\theta ) $ and
$\Theta = m^{1+\varepsilon} \theta$. This gives
\begin{eqnarray}
\label{eq:perturb-flow-rescaled}
-m \partial_m  w(\theta)&=& (2+\varepsilon) w(\theta) - (1+\varepsilon)
  \theta w'(\theta)  \nonumber \\
  && +\, 2 \left( \theta w'(\theta)w''(\theta)  - \tilde{N}  w'(\theta)^2 \right ).
\end{eqnarray}
The $U(N)$ GN model corresponds to the model~(\ref{eq:action-Weyl-replicated}) with $W(\Theta)$
being a quadratic function, so that the  FP of the GN model can be easily identified
with $w^*(\theta)=\varepsilon \theta^2/[8(1- \tilde{N})] $.
If we restrict $w(\theta)$ to a quadratic function
its amplitude remains the only  unstable direction.
To check the full stability  we
linearize the flow~(\ref{eq:perturb-flow-rescaled})
around this FP.
The derivatives of the characteristic function $w^{(n)}(0)$ are
coupled to the operators  $\Theta^n$ corresponding
to the fermion density moments. Using~(\ref{eq:perturb-flow-rescaled}) we
can calculate their scaling dimensions
$[w^{(n)}(0)] = 2+\varepsilon - n(1+\varepsilon) + \varepsilon n (2\tilde{N}-n) /(2\tilde{N}-2)$
which are in agreement with diagrammatic~\cite{Louvet:2016} and conformal
	field theory~\cite{Ghosh:2016} results.
The coupling is relevant if its scaling dimension is positive.
Hence in the limit of $N\to 0$, which describes the disordered relativistic
semimetal, infinitely many relevant operators corresponding to higher order
cumulants of the disorder distribution  are identified signaling the
relevance of rare configurations of disorder at the
transition~\footnote{Note that in the limit of large $N$, the  $\theta^2$ remains
the only relevant operator, and thus,  the GN FP does describe the transition.
In this limit  Eq.~(\ref{eq:perturb-unrescaled}) can be transformed
into  the inviscid Burgers equation, which develops a shock at the origin
related to the fermion mass generation  (see Supplemental Material~\cite{Supplementary}). }.

\textit{Zero $N$ limit and porous medium equation. --}
Since in the limit of $N \to 0$ the GN FP  becomes unstable in
infinitely  many  directions,  it cannot control a continuous  transition.
Nevertheless, it is premature to conclude that the transition is smeared out
or first order.
A direct numerical integration of the rescaled flow
equation~(\ref{eq:perturb-flow-rescaled}), however,  failed to find any physical FP
different from the GN one. As we will see below, this can be explained
by the fact that the FP we are looking for has a nonanalytical behavior at the origin
in addition to the absence of boundary conditions at infinity.
Notice, however, that if the large $\theta$ asymptotics  of the FP $w^*(\theta)$
was known then the whole FP could be computed by numerical
integration of Eq.~(\ref{eq:perturb-flow-rescaled}).
Fortunately, introducing
the \textquotedblleft time\textquotedblright $t=(m^{\varepsilon}_0-m^{\varepsilon})/\varepsilon$,
the \textquotedblleft coordinate\textquotedblright $r=\sqrt{2\Theta}$ and
the \textquotedblleft density profile\textquotedblright  $u= W'(\Theta)$ we can
rewrite the unrescaled flow equation~(\ref{eq:perturb-unrescaled}) in the form
of a 2D nonlinear diffusion equation
\begin{eqnarray}
\label{eq:PME}
2 \partial_t u(r,t)=\frac1{r}\partial_r r\partial_r u^2(r,t) = \Delta u^2(r,t)
\end{eqnarray}
with the superimposed radial symmetry. Since $m$ changes from $m_0$ to $0$
one has to stop the evolution of the density profile $u(r,t)$ at the maximal
observation time $T_0= m^{\varepsilon}_0/\varepsilon$.
Equation~(\ref{eq:PME}) is the 2D
PME which has been intensively studied by mathematicians for  several
decades~\cite{Vazquez:2007}. Because of
the presence of degeneracy points (regions where $u=0$ and thus vanishing  diffusion constant)
the PME exhibits remarkable nonlinear phenomena. They include
finite velocity propagation of fronts separating the regions with zero and
nonzero~$u$~\cite{Gratton:1990},
waiting times before the front starts to move~\cite{Giacomelli:2006},
and self-focusing solutions describing shrinking of holes
in the support of~$u$~\cite{Aronson:2003} with postfocusing  accumulation of
diffusing particles~\cite{Angenent:1996}.
Following the route paved by these mathematical studies
we look for a backward self-similar
solution (BSS) to Eq.~(\ref{eq:PME}) which has the form
\begin{equation}
\label{eq:self-similar-sol}
u(r,t) = (T-t)^{2\delta-1}F(\zeta), \ \ \ \ \zeta = \frac{r}{(T-t)^\delta}.
\end{equation}
The self-similar solutions to the PME  play a special role since they
lead to a universal large time behavior.
It is straightforward to identify  the BSS~(\ref{eq:self-similar-sol}) with a FP
solution $w(\theta)$ to the rescaled FRG equation~(\ref{eq:perturb-flow-rescaled})
setting %with
$\delta=(1+\varepsilon)/(2\varepsilon)$, $T=T_0$ and $F(\zeta) = \varepsilon
 w' \left(\zeta^2/(2\varepsilon^2) \right)$.
Then the rescaled FRG equation~(\ref{eq:perturb-flow-rescaled}) becomes
\begin{eqnarray}
\label{eq:for-F}
\partial_{\tau}F(\zeta) &=& F(\zeta)\left[ F''(\zeta) +\frac1{\zeta} F'(\zeta)\right]
+ F'(\zeta)^2 \nonumber \\
&& -\delta \zeta F'(\zeta)  +(2\delta-1) F(\zeta),
\end{eqnarray}
where we have defined $\tau=-\ln(T-t)$ such that
$\varepsilon \partial_\tau = - m \partial_m $.
%%%%%%%%%%%%%%%%%%%%%%%%%%%%%%%%%%%%%%%%%%%%%%%%%%%%%%%%%%%%%%%%%%%%%%%%%%%%%%%%%%%%%%%%%%%%%%%%%%%%
\begin{figure}
\includegraphics[width=8.5cm]{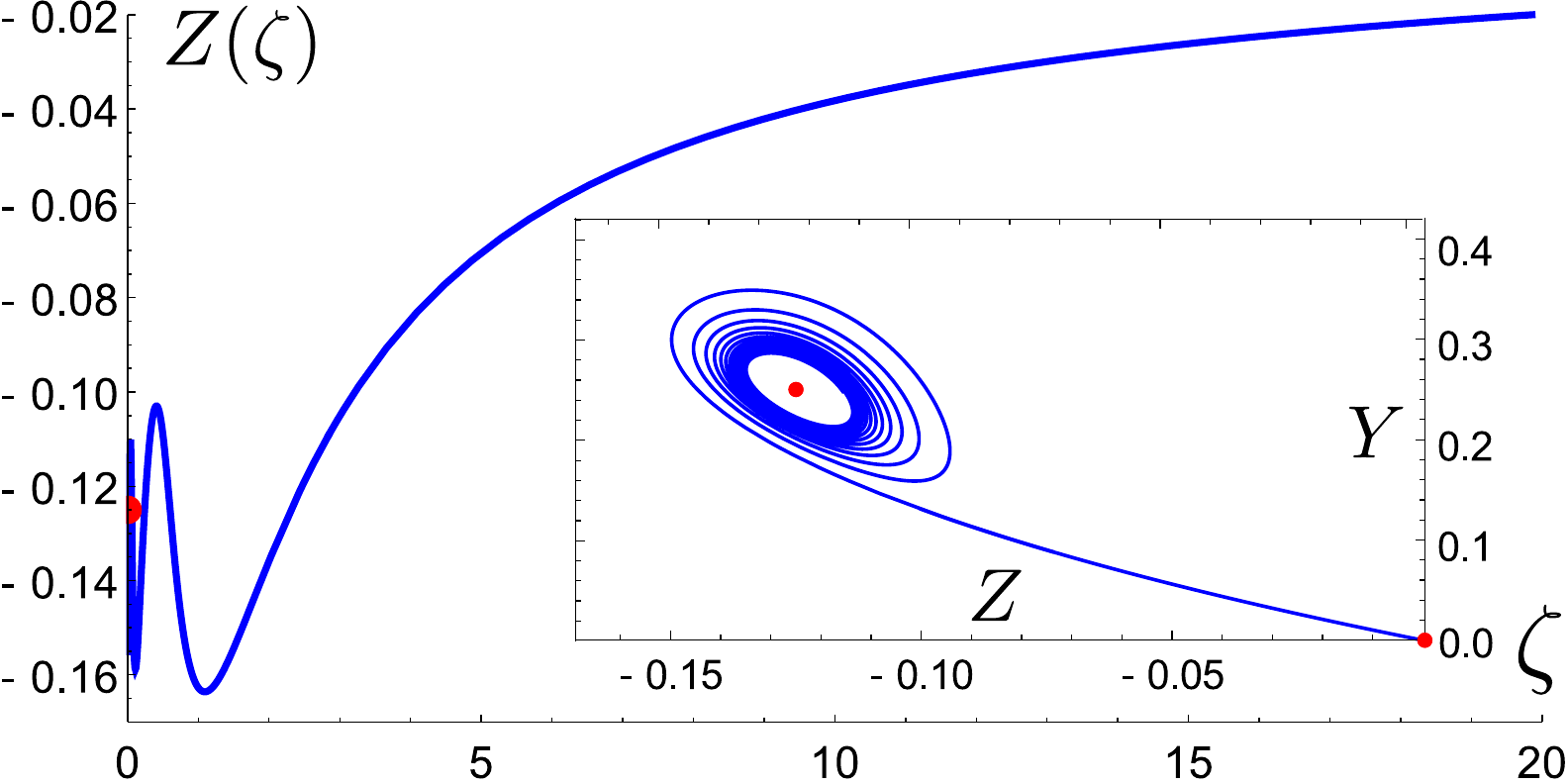}
 \caption{
 Fixed point solution to  the  functional renormalization group
 equation for $d=3$ expressed as a backward self-similar solution $Z(\zeta)$
 to the porous medium equation~\eqref{eq:PME}
 with $\delta=1$.
 The inset shows the integral curve representing the BSS
 in the phase plane $(Z,Y)$, which clarifies the nature of nonanalyticity of the FP.
 }
 \label{fig:Z}
\end{figure}
%%%%%%%%%%%%%%%%%%%%%%%%%%%%%%%%%%%%%%%%%%%%%%%%%%%%%%%%%%%%%%%%%%%%%%%%%%%%%%%%%%%%%%%%%%%%%%%%%%
For a BSS $F(\zeta)$ the rhs of Eq.~(\ref{eq:for-F}) identically vanishes.
The  GN FP  corresponds to the BSS with  $F(\zeta)=\zeta^2/8$.
One may get the impression that  rewriting the FRG equation~(\ref{eq:perturb-flow-rescaled})
in the  form~(\ref{eq:for-F}) is just a beautiful mathematical trick
which connects two \textit{a priori} unrelated problems.  However, there is much more
to it than that.
Indeed, while the BSS~(\ref{eq:self-similar-sol}) translates into a
FP  at $T=T_0$,
as we will see below, its dependance on $T$
also provides an explicit expression for the flow of the
whole disorder distribution along a single unstable direction.
Moreover, the nontrivial BSS~(\ref{eq:self-similar-sol})
can be captured
by the phase-plane formalism~\cite{Vazquez:2007} which is a powerful tool
for  analysis of the PME~(\ref{eq:PME}).
To that end  we define the phase variables $Z$ and $Y$  as
$F(\zeta)=-\zeta^2 Z(\zeta)$ and $Y(\zeta)=-(2Z(\zeta)+\zeta Z'(\zeta))$.
They satisfy  autonomous first order differential
equations~\cite{Supplementary}
whose solution for $\delta=1$, i.e., $d=3$,
is shown in Fig~\ref{fig:Z}. In the phase plane $(Z,Y)$  the BSS is
represented by an integral curve which connects
the singular point $(0,0)$ controlling the large $\zeta$ behavior
and a limiting cycle
around the singular point $(-\frac18,\frac14)$ corresponding to the
GN FP (see inset of Fig~\ref{fig:Z}).
Although the
function $Z(\zeta)$ is infinitely oscillating at the origin,  the corresponding
profile function $F(\zeta)$ is surprisingly monotonic as one can see
in Fig.~\ref{fig:F}.
It grows as $F(\zeta) \sim \zeta^{2-1/\delta}$
for large $\zeta$ and is strongly  nonanalytic  at  $\zeta=0$.
This explains why the nontrivial FP can be easily
overlooked when solving numerically the FRG equation~(\ref{eq:perturb-flow-rescaled}).
The new nonanalytic  FP exists only for
$\delta>\delta_c \approx 0.856\,326\,5$, i.e.
only below the critical dimension $d_{c} \approx 3.4$, and thus  controls
the transition in $d=3$.

\begin{figure}
\includegraphics[width=8.5cm]{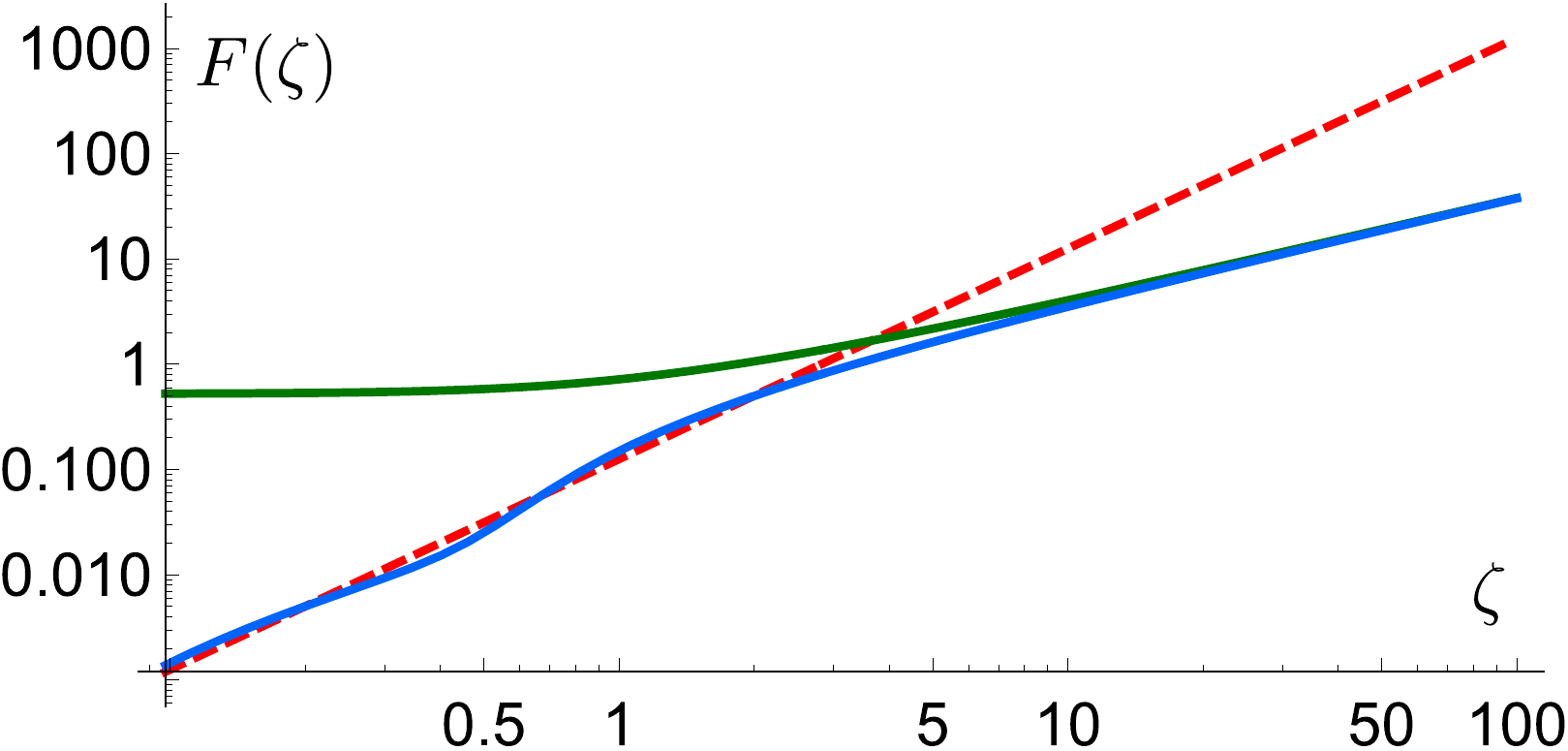}
 \caption{ Analytical continuation of the backward self-similar
 solution~(\ref{eq:self-similar-sol}) for $t<T$ to the forward
 self-similar solution~(\ref{eq:post-focus}) for $t>T$ which
 provides a nonanalytic mechanism for the DOS generation
 at the nodal point. The generated DOS is related to the nonzero $\tilde{F}(0)$.
 Blue solid line is the BSS  function $F(\zeta)$
 for $\delta=1$, green solid line is the corresponding FSS
 $\tilde{F}(\tilde{\zeta})$  with $\tilde{F}(0)>0$. Both solutions have the same
 asymptotic behavior at large $\zeta$,
 which ensures the matching between $u(r,T^{-})$  and $u(r,T^{+})$.
 Red dashed line is
 the  GN FP $F(\zeta) = \zeta^2/8$ shown for comparison.
  }
 \label{fig:F}
\end{figure}
%%%%%%%%%%%%%%%%%%%%%%%%%%%%%%%%%%%%%%%%%%%%%%%%%%%%%%%%%%%%%%%%%%%%%%%%%%%%%%%%%%%%%%%%%%%%%%%%%%

\textit{Stability analysis. --}  To study the stability of the new nonanalytic FP
we add to the BSS~(\ref{eq:self-similar-sol}) a time dependent
perturbation
$F(\zeta;\tau) =  F(\zeta)+ \phi(\zeta) e^{\lambda\tau}$, where $\lambda$
is the stability
eigenvalue and $\phi$ is the corresponding eigenfunction.
Substituting it into  Eq.~(\ref{eq:for-F}) and linearizing around the
BSS we arrive at
\begin{eqnarray}
\label{eq:stability}
- Z \ddot{\phi} + \left[2 Y  - \delta  \right] \dot{\phi}
+ \left[2\delta-1 -\lambda +  2Y+  \dot{Y} \right] \phi = 0,
\end{eqnarray}
where the dots stand for the logarithmic derivatives, $\dot{X} \equiv d X/ d\ln \zeta $.
In order to obtain the stability spectrum of the FRG FP
one has to impose the boundary condition at $\zeta=0$ using additional physical arguments~\cite{Wiese:2016}.
Here we look for  perturbations originating from
higher order cumulants. Choosing $\phi_n(\zeta) = \zeta^n f_n(\zeta)$,  $n = 2,4,6,...$ such that the functions
$f_n(\zeta)$ are bounded for $\zeta \to 0$ but not necessarily analytic we render
the spectrum discrete.
Numerical solution of Eq.~(\ref{eq:stability}) shows that
only the eigenvalue corresponding to
$n = 2$ is positive, and thus, the nonanalytic FP we have found indeed is a
critical FP describing the disorder driven transition~\cite{Supplementary}.

Remarkably, the relevant eigenvalue and eigenfunction can be identified from general symmetry considerations.
Let $u(r,t)$ be a BSS with profile $F(\zeta)$ and  waiting time $T$.
Owing to the time-translational
invariance of the PME~(\ref{eq:PME}) we can shift
$T \to T+\Delta T$ to obtain another BSS:
$u(r,t) \to u(r,t) + \Delta T \partial_T u(r,t)$.
From
Eq.~(\ref{eq:self-similar-sol}) we find that
$\partial_T u(r,t) =  (T-t)^{2\delta-1} \phi_2(\zeta) e^{\tau}$ where
\begin{eqnarray}
\label{eq:eigenfunction}
\phi_2(\zeta) = (2\delta-1) F(\zeta)  -\delta \zeta  F'(\zeta).
\end{eqnarray}
It is straightforward to see that~(\ref{eq:eigenfunction}) is the eigenfunction of
Eq.~(\ref{eq:stability}) which corresponds to the only positive eigenvalue $\lambda_2=1$.
Thus, $T_0-T$ is the only relevant parameter
which controls the transition.
Taking into account the relation between $\tau$ and $m$  we can find the
correlation length exponent $\nu^{-1} = \varepsilon \lambda_2 $.
Although the nonanalytic FP can always be  expressed as a
BSS~(\ref{eq:self-similar-sol}) with $T=T_0$,  higher loop order corrections to
the critical exponents are expected.

\textit{Postfocusing regime and DOS generation. --}
The inverse waiting time $T^{-1}$ determined by the full bare
disorder distribution
\footnote{There is no simple way to compute the waiting time for
the PME from
the initial profile function $u(r,t=0)=u_0(r)$ determined by the bare disorder
distribution. However, it existence as well as the upper and lower
bounds have been proved for some cases.
As an illustration, it was shown in~\cite{Giacomelli:2006} that
for the initial condition $u_0$ such that
$A r^2  \le  u_0(r) \le B r^2 $  the waiting time
satisfies the inequalities  $1/(2B) \le T  \le 1/(2A)$.}
turns out to be a natural measure of the disorder strength.
If the bare disorder is weak ($T>T_0$)  the system is in the semimetal
phase, while for strong disorder ($T<T_0$) it is
in the diffusive phase. The disorder is critical for $T=T_0$.
In order see how the DOS at the
zero energy is generated by the FRG flow
the BSS~(\ref{eq:self-similar-sol}) of PME~(\ref{eq:PME}) corresponding to $T<T_0$  has to be continued
analytically from $t<T$ to $t>T$.
Recalling  the  asymptotic  behavior $F(\zeta) \sim \zeta^{2-1/\delta}$ for $\zeta \to \infty$
we find by the continuity that $u(r,t=T) = c^* r^{(2\delta-1)/\delta}(1+o(1))$ for
$r \to 0$. In the postfocusing regime, i.e. for  $t>T$, the fictional
particles, whose nonlinear
diffusion is described
by the PME~(\ref{eq:PME}), start to accumulate at the origin. This is
described by a forward
self-similar solution (FSS)~\cite{Angenent:1996}:
\begin{equation}
 \label{eq:post-focus}
 u(r,t)=(t-T)^{2\delta-1}\tilde{F}(\tilde{\zeta}),
 \ \ \ \tilde{\zeta}=\frac{r}{(t-T)^\delta}.
\end{equation}
The FSS~(\ref{eq:post-focus}) can be found using the same phase-plane
formalism~\cite{Supplementary}.
It implies that
$\tilde{F}(0)= \mathrm{const} $ and
$\tilde{F}(\tilde{\zeta}) \sim \tilde{\zeta}^{2-1/\delta}$ for
$\tilde{\zeta} \to \infty$ (see Fig.~\ref{fig:F}).
Since $W'(0^+) = u(0,t) \neq 0$ in the postfocusing regime,
the FSS~(\ref{eq:post-focus}) describes the diffusive phase of
relativistic fermions and allows one to compute explicitly the DOS
at zero energy. We find that
close to the transition, i.e. for $T_0-T \ll T_0$
the DOS at zero energy is given by $\overline{\rho(0)}\sim(T_0-T)^{\beta} $ with
the order parameter critical exponent $\beta = 2\delta-1$.
Assuming that the hyperscaling relation $\beta=(d-z)\nu$  is not broken
we obtain the dynamic critical exponent as $z=1+\varepsilon+O(\varepsilon^2)$.
%%%%%
Beside the averaged DOS
the postfocusing regime of the FRG flow~(\ref{eq:post-focus})
allows us to characterize the scaling behavior of the whole distribution of its fluctuations
in the diffusive metal phase.
We find that the scaling behavior of the $n$th cumulant of the DOS fluctuations  at zero energy scales as
$\overline{\rho^n(0)}^{c}\sim(T_0-T)^{\beta-2\delta(n-1)} $ close to the transition.
This scaling signals that the corresponding
distribution becomes very broad when one approaches the transition from
the diffusive phase.

\textit{Conclusions and outlook. --}
We have developed a FRG approach to
the semimetal-diffusive metal transition in disordered Weyl
fermions. We have shown that the previously studied
FP corresponding to the Gaussian distribution of disorder
is unstable, demonstrating the relevance of rare disorder fluctuations
at this transition.
Indeed, the analysis of the flow equation derived in
a fixed dimension $d$ reveals the proliferation
of infinite number of higher order cumulants in the
running disorder distribution, even if starting from a pure
Gaussian distribution~\cite{Supplementary}.
In order to resolve this problem we have established a connection between
the FRG equation and the celebrated PME, whose self-similar solution
represents a nonanalytic FP describing the transition.
Its analytical continuation to the postfocusing  regime
provides a unique mechanism of spontaneous generation of a finite
DOS at zero energy
\footnote{This mechanism is remarkably different from that of the GN model in the limit of large $N$
given by the mean-field theory (see e.g. D. D. Scherer, J. Braun, and H. Gies,
\href{https://doi.org/10.1088/1751-8113/46/28/285002}{J. Phys. A: Math. Theor. \textbf{46}, 285002 (2013)})}.
Moreover, it shows that the distribution of fluctuations
becomes very broad close to the transition.
In particular, one expects that the
critical wave functions  exhibit multifractality at the transition~\cite{Pixley:2015},
with a spectrum different from that at the GN FP~\cite{Louvet:2016,Syzranov:2016a}.

It was argued that rare
disorder configurations can give rise to
a finite DOS in the semimetal phase~\cite{Nandkishore:2014}.
Although more refine recent calculations~\cite{Buchhold:2018}
suggest that their contribution
vanishes in the thermodynamic limit, a finite DOS was observed in numerical
simulations~\cite{Pixley:2016}. While this can be due to a simple finite size
correction,  $\overline{\rho(0)} \sim L^{z-d}$,  our FRG description
provides another mechanism. Indeed, vanishing of the DOS
stems from the existence of a finite waiting time in
the FRG flow~\eqref{eq:PME}. Once one introduces
disorder correlations similar to that used in Ref.~\cite{Nandkishore:2014}, the waiting
time phenomenon is then superimposed  with a slow creeplike motion, which
generates a finite DOS. Nevertheless the universal critical properties
of the underlying FP will dominate over the nonuniversal contributions
depending on the UV cutoff.

Beyond the present transition, our approach can be applied to other problems where nonanalyticity
in the renormalized disorder distribution may
play a significant role, such as the Anderson localization transition.
In that case,  the relevance of infinitely many so-called high-gradient operators
at the  conventional
FP of the corresponding nonlinear sigma
model~\cite{Kravtsov:1989,Wegner:1990,Castilla:1993}
raises the question of the existence
of a new FP which indeed controls the transition.
An argument against such an unconventional FP~\cite{Wegner:1990}
can also be applied to the present semimetal - diffusive metal transition.
Here, however, it is ruled out by nonanalyticity of the new FP, which hints at a similar scenario
in the case of the Anderson localization
transition~\footnote{In~\cite{Wegner:1990} it was argued that
fusion of two  (high-gradient) operators of the same rank $n$ yields
a  contribution of rank $2n-2$, the feedback to the same rank operator
is missing, and thus no FP  different from the conventional one can be obtained.
Exactly the same argument is applied
to Eq.~(\ref{eq:perturb-flow-rescaled}) rewritten as a
system of flow equations
for $w^{(n)}(0)$, but it is invalidated by nonanalyticity of the new FP}.

\begin{acknowledgments}
\textit{Acknowledgments. --} We would like to thank
Krzysztof Gaw\c{e}dzki,  Fabio Franchini and Pierre Le Doussal
for inspiring discussions and
critical reading of the manuscript.
I.B. acknowledges the support of the Croatian
Science Foundation Project No. IP-2016-6-3347 and the QuantiXLie Centre of Excellence, a project
cofinanced by the Croatian Government and European Union through the
European Regional Development Fund - the Competitiveness and Cohesion
Operational Programme (Grant KK.01.1.1.01.0004).
D.C. and A.A.F. acknowledge the support from the French Agence Nationale de
la Recherche by the Grant No. ANR-17-CE30-0023 (DIRAC3D).
\end{acknowledgments}


%%%%%%%%%%%%%%%%%%%%%%%%%%%%%%%%%%%%%%%%%%%%%%%%%%%%%%%%%%%%%

%\bibliographystyle{apsrev4-1}
%\bibliography{disorderedWeyl}

%%%%%%%%%%%%%%%%%%%%%%%%%%%%%%%%%%%%%%%%%%%%%%%%%%%%%%%%%%%%%

%%%%%%%%%%%%%%%%%%%%%%%%%%%%%%%%%%%%%%%%%%%%%%%%%%%%%%%%%%%%%

\pagebreak \newpage

\widetext

\begin{center}
\Large \bf
  SUPPLEMENTAL MATERIAL
\end{center}

\newcounter{CounterShift}
\setcounter{CounterShift}{\value{equation}}
\renewcommand{\theequation}{S.{\the\numexpr\value{equation}-\the\numexpr\value{CounterShift}\relax}}

The Supplemental Material is organized as follows.
In Sec.~\hyperlink{sec:A}{A} we derive the Nonperturbative functional RG (NPRG) equation directly in $d$ dimensions.
Using these results in Sec.~\hyperlink{sec:B}{B} we obtain the flow equation which
is perturbative in $\varepsilon=d-2$.
In Sec.~\hyperlink{sec:C}{C} we show how the FRG flow equation can be mapped
onto the inviscid Burgers equation in the limit of large $N$.
In Sec.~\hyperlink{sec:D}{D} we explain the difference between
the classical and weak (or integral) solutions. In Sec.~\hyperlink{sec:E}{E} we give the details
of the phase-plane analysis of the PME which allows us to find a BSS corresponding to a nonanalytic FP of the
FRG equation. Section~\hyperlink{sec:F}{F} contains details of the stability analysis. In
Sec.~\hyperlink{sec:G}{G} the phase-plane formalism is used for studying the postfocusing regime which describes
generation of the DOS at zero energy in the diffusive phase.

\hypertarget{sec:A}{}

\section{A. Nonperturbative Functional RG approach}\label{sec:NPRG}

Here we give an overview of the derivation of the RG
flow equation for the characteristic function starting from the Wetterich effective
average action approach~\cite{Berges:2002}. All the notions are fairly standard and technical and more
details can be found in~\cite{Hofling:2002,Rosa:2001} where the account is given of the
GN model, which is formally very similar to the problem we study. One has to define
the generating functional using the replicated bare level action~(2), with an addition
of the infrared regulator~\cite{Mouhanna:2010}
\begin{equation}
\label{reg_general}
  \Delta S_k[\bar{\psi},\psi]=\int d^dx d^dy \bar{\psi}^i_{\alpha}(x) R_{k;\alpha,
   \beta}^{ij}(x-y)\psi_\beta^j(y),
\end{equation}
which includes $k$ as the cutoff wave vector and $\alpha$ and $\beta$ are replica indices.
The regulator modifies the bare action by decoupling the fast modes ($q>k$) from the slow ($q<k$) ones
by giving a large mass to the latter and leaving the first unaffected. With the bare action
modified in this way and with the addition of external sources, one can construct a $k$ dependent generating functional as a path integral
\begin{equation}
  Z_k[\bar{B},B]=\int \mathcal{D}\bar{\psi} \mathcal{D} \psi\,
  e^{-S-\Delta S_k+\int d^d x \bar{\psi}(x)B(x)+\int d^d x
  \bar{B}(x) \psi(x)},
\end{equation}
where we used a shortcut notation $\bar{\psi} \psi := \sum_{i,\alpha} \bar{\psi}_{\alpha}^i\psi_{\alpha}^i$, and similarly for the terms with external sources.

To define the effective average action $\Gamma_k$, explicitly dependent on the cutoff $k$, one
performs the Legendre transform on the logarithm of the generating functional
\begin{equation} \label{eq:Legendre-transform}
  \Gamma_k[\bar{\Psi},\Psi]+\ln(Z_k[\bar{B},B])=\int d^dx \bar{\Psi}(x)B(x)
  +\int d^dx \bar{B}(x)\Psi(x)+\Delta S_k[\bar{\Psi},\Psi]
\end{equation}
where $\Psi=\langle\psi\rangle$ and $\bar{\Psi}=\langle\bar{\psi}\rangle$ are formal ensemble averages
of $\psi$ and $\bar{\psi}$, respectively.

To derive the exact flow equation for  the effective action~(\ref{eq:Legendre-transform})  we first
combine the two fermion fields in a 4-component spinor $(\bar{\Psi}_a,\Psi_a)$ and then
construct the regulated second derivatives matrix $\Gamma^{(2)}$ as~\cite{Berges:2002}
\begin{equation} \label{eq:4-component-propagator}
  \Gamma^{(2)}_{\alpha,\beta}(x,y)+R_{k;\alpha,\beta}(x-y)=\Bigg( \begin{array}{cc}
\frac{\overrightarrow{\delta}}{\delta\bar{\Psi}_{\alpha}(x)}(\Gamma+\Delta S_k)\frac{\overleftarrow{\delta}}{\delta\bar{\Psi}_{\beta}(y)} & \frac{\overrightarrow{\delta}}{\delta\bar{\Psi}_{\alpha}(x)}(\Gamma+\Delta S_k)\frac{\overleftarrow{\delta}}{\delta\Psi_{\beta}(y)}\\
\frac{\overrightarrow{\delta}}{\delta\Psi_{\alpha}(x)}(\Gamma+\Delta S_k)\frac{\overleftarrow{\delta}}{\delta\bar{\Psi}_{\beta}(y)} & \frac{\overrightarrow{\delta}}{\delta\Psi_{\alpha}(x)}(\Gamma+\Delta S_k)\frac{\overleftarrow{\delta}}{\delta\Psi_{\beta}(y)} \end{array}\Bigg).
\end{equation}
Here we have omitted the spinor indices and used the notation of the left and right derivatives,
which is useful when considering the action depending on Grassmann (anticommuting) variables.
We calculate  the  "full"  $\bar{\Psi}$,$\Psi$-dependant exact propagator
by inverting the expression (\ref{eq:4-component-propagator}) as
\begin{equation} \label{eq:full-propagator}
  \mathcal{G}_{k;\alpha,\beta}(x-y)=(\Gamma^{(2)}+R_{k})^{-1}_{\alpha,\beta}(x-y).
\end{equation}
The exact flow equation for the effective action can be written as~\cite{Berges:2002}
\begin{equation}
\label{eq:Wetterich}
  \partial_s\Gamma_s=-\frac12  {\rm Tr }\Bigg\{ (\partial_s R_{k})
  \mathcal{G}_{k}\Bigg\},
\end{equation}
where the symbol $\rm Tr$ stands for the trace over all indices and integration over space.
In Eq.~(\ref{eq:Wetterich}) we have also introduced the RG time $s=\ln(k/\Lambda)$ going from $0$
for the microscopic
theory to $-\infty$ for a theory where all the degrees of freedom have been integrated out.
Here $\Lambda$ is the UV cutoff, e.g. the width of the  Brillouin zone.
The minus in Eq.~(\ref{eq:Wetterich}) is the usual extra minus associated
with a fermionic loop.

To make the flow equation for $\Gamma$ tractable, one needs an approximation scheme for the
effective average action, since the flow equation~(\ref{eq:Wetterich}) can not be solved exactly.
The most simple approximation is the so-called local potential approximation (LPA) in which
one takes into account the nonlocal space dependence only in the gradient term of the single replica term in
the bare action~(2). It is known that already this approximation is
able to properly capture the long-wavelength scaling behavior of many systems~\cite{Delamotte:2012}.
We start from the following ansatz for the effective average action
\begin{equation}
\label{eq:LPA-ansatz}
  \Gamma_k =\int d^dx\,  i \bar{\Psi}_{\alpha}(x) (-i \vec{\gamma}\vec{\partial}-i\omega)
   \Psi_{\alpha}(x) + W_k [\Theta(x)].
\end{equation}
where  $\Theta(x)$ is the local density as in Eq.~(2) of the main paper and $W_k$ is
the renormalized characteristic function.

After a tedious, but straightforward derivation, which will be detailed elsewhere,
one arrives at the following RG flow equation for the characteristic function
\begin{eqnarray}
 \partial_s W_s = -\frac{1}{2}  \int_q \partial_s R_k(q)
 \left\{ \frac{2(\tilde{q}\cdot q) N \mathrm{tr}\mathbb{I} }{\tilde{q}^2+B^2} +
  \frac{4  B (\tilde{q}\cdot q) \Theta W''
  (\Theta)}{(\tilde{q}^2+B^2)[\tilde{q}^2+B^2 +2B\Theta W''(\Theta)]} \right\}.
  \label{eq:flow-1}
\end{eqnarray}
In Eq. (\ref{eq:flow-1}), $B=\omega+W'(\Theta)$, $\mathbb{I}$ is the  unity matrix in the spinor space
(e.g. for a single 3D Weyl cone  $\mathrm{tr} \mathbb{I}=2$) and $\tilde{q}$ is the momentum modified
by the regulator, defined in a similar way as in e.g. Ref.~\cite{Delamotte:2012}. It is interesting to note
that the equation similar to  Eq.~({eq:flow-1}) has been derived in~\cite{Jakovac:2013}
but in a different context. It is easy to see from  Eq.~({eq:flow-1})
that even if the bare disorder has a pure Gaussian distribution with $W(\Theta) \sim \Theta^2$
the higher order cumulants will be ultimately generated by the FRG flow due to the non-trivial
denominator in the second term of Eq.~(\ref{eq:flow-1}).

\hypertarget{sec:B}{}
\section{B. Perturbative version of the flow equation} \label{sec:perturbative}

Analysis of Eq.~(\ref{eq:flow-1}) is rather complicated. In order to gain the physics insight
we will restrict ourselves to functional but perturbative in $\varepsilon=d-2$ RG equation. In principal
this equation can derived by computing the diagrams similar to those introduced in
Supplementary material of paper~\cite{Louvet:2016}.
However, instead of direct computing diagrams for a field theory with infinitely many coupling constants
represented by the high order cumulants
one can use the NPRG flow equation~(\ref{eq:flow-1}) as a starting point
to derive the one-loop  perturbative FRG equation in $d=2+\varepsilon$.
To that end we rewrite  Eq.~(\ref{eq:flow-1}) in an infinitesimal form replacing the
cutoff function $\partial_s R_k(q)$  by imposing dimensional regularization~\cite{Zinn-Justin:1986}.
This gives the one-loop correction to the characteristic function
\begin{eqnarray}
 \delta W(\Theta) = 2  \int_q
 \left\{\tilde{N} - \frac{ W'(\Theta)^2 \tilde{N} }{{q}^2+B^2} +
  \frac{  q^2 \Theta W'(\Theta)W''(\Theta)
  }{({q}^2+B^2)[{q}^2+B^2 +2B\Theta W''(\Theta)]} \right\}.
  \label{eq:flow-3}
\end{eqnarray}
We now use the dimension regularization properties in order to simplify
Eq.~(\ref{eq:flow-3}). Since $\int_q 1 =0 $ in dimensional regularization, the first term in r.h.s.
of Eq.~(\ref{eq:flow-3}) vanishes.
To one-loop order only the poles in $\varepsilon$ should be kept in the rest two terms. We find
\begin{eqnarray}
 \int_q  \frac{1}{{q}^2+B^2}  = - S_d\frac{B^{\varepsilon}}{\varepsilon} + O(\varepsilon).
  \label{eq:DR-integral-1}
\end{eqnarray}
and
\begin{eqnarray}
 \int_q  \frac{q^2}{({q}^2+B^2)({q}^2+B^2+\mathrm{const} B)}
 = -S_d \frac{B^{\varepsilon}}{\varepsilon} + O(1),
  \label{eq:DR-integral-2}
\end{eqnarray}
where $S_d$ is the area of $d$ -dimensional sphere divided by $(2\pi)^d$.
It is easy to see that $B$ plays the role of IR cutoff in Eqs.~(\ref{eq:DR-integral-1}) and
(\ref{eq:DR-integral-2}). Replacing it by mass $m$ for convenience and collecting all the terms we arrive at
\begin{eqnarray}
 \delta W(\Theta) = -2S_d\frac{m^{\varepsilon}}{\varepsilon}
  \left[ \Theta W'(\Theta)W''(\Theta) -  W'(\Theta)^2 \tilde{N} \right].
  \label{eq:flow-4}
\end{eqnarray}
One can now identify the terms in Eq.~(\ref{eq:flow-4}) with the contributions coming from the
one-loop diagrams in  the perturbative loop expansion.
To compute the one-loop perturbative FRG flow equation we follow Refs.~\cite{LeDoussal:2004,Fedorenko:2006}
and define the renormalized characteristic function as
$  W_R(\Theta) = W(\Theta) +\delta W(\Theta)$.
Taking a derivative $-m\partial_m$
on both sides and reexpressing the obtained equation in terms of the renormalized characteristic
function $W_R$ we obtain the flow equation~(3)
from the main paper (after
including $S_d$ into redefinition of $W_R$).

\hypertarget{sec:C}{}
\section{C. Large $N$ limit and Burgers equation} \label{sec:Burgers}

Here we discuss how the conventional GN FP describes  the chiral
transition of the  $N$-flavor
interacting relativistic massless fermions in the limit of large $N$.
Since this transition is driven by  repulsive interactions, the GN FP $w^*$ is negative in our convention.
We find that for $d>2$
the only relevant coupling constant is the usual strength of fermion repulsion $-w''(0)$
which controls the transition. The corresponding positive eigenvalue gives the correlation length exponent
$\nu$.
The nonanalytic nature of the transition can be revealed by introducing $y(r)=W'(\Theta)$,
$\Theta= 4\tilde{N} r$ and the "time" $t=(m^{\varepsilon}_0-m^{\varepsilon})/\varepsilon$,
directly in the unrescaled flow equation~(3) in the main text.
Then, in the limit $\tilde{N} \to \infty$,  it transforms into the inviscid Burgers
equation~\cite{Bec:2007}
\begin{eqnarray}
\label{eq:Burgers}
\partial_t y(r) +  y(r)y'(r) =  0,
\end{eqnarray}
which can be solved by the method of characteristics. The reasonable bare interactions between
fermions correspond to the initial condition with an odd function $y_0(r)$ satisfying $y_0(0)=0$.
After  waiting (breaking) time $T^* =  1/ |y'_0(0)|$
the "velocity" profile $y(r)$ develops a shock exactly at the origin $r=0$.
The appearance  of the discontinuity  $y(0^+)=-y(0^-)>0$ preserves the single-value property of
the classical solution and transforms it into the so-called weak (or integral)
solution~(see Section~\hyperlink{sec:D}{D}).
The shock leads to $W'(0^+) \neq 0$, and thus,
to the dynamical fermion mass generation and spontaneous  chiral symmetry breaking.
However, since $0\le m\le m_0$ one has to
stop the evolution of the profile $y(r)$ at the maximal
observation time $T_0= m^{\varepsilon}_0/\varepsilon$.
Thus, we conclude that the system is in the symmetric phase
if the bare interactions are so small that the waiting time  $T^*$
is larger than the maximal observation time $T_0$. If $T^*<T_0$ then the system is
in the  symmetry broken phase and the shock size at final time
$T_0$ gives the value of the order  parameter.
A similar mechanism was recently proposed for the chiral symmetry breaking
transition in the 4D Nambu-Jona-Lasinio model~\cite{Aoki:2014}.

\hypertarget{sec:D}{}
\section{D. Weak  vs. strong solution} \label{sec:weak}

The important physical implications obtained from the  flow
equation for the characteristic function $W$ detailed in the main text,
rely on the spontaneous generation of a cusp at the origin.
This is possible only if we extend the definition of solution to the FRG flow equation
from the classical strong solution to the so-called ``weak solution"~\cite{Evans:1998}.
Contrary to the strong solution a weak solution may contain discontinuities and may not be differentiable.
To define a weak solution to a PDE one usually needs to reformulate the problem in an integral form.

As an example, let us consider the Burgers equation~(\ref{eq:Burgers})
which has a form of a conservation law.
The solution to the first order PDE~(\ref{eq:Burgers})
constructed by  the method of characteristics
becomes a multivalued function beyond the so-called breaking time. The multivalued parts
can be eliminated by inserting shocks using the equal area rule which is a result of conservation.
In other words the integral of the discontinuous weak
solution with shock must be the same as the integral of the auxiliary multivalued solution.
In order to avoid the necessity
of using the equal area rule let us integrate the Burgers equation~(\ref{eq:Burgers}) convoluted with a
smooth bounded test function $\phi(r, t)$. Then, after integrating  by parts we arrive at
\begin{eqnarray}
\label{eq:weak-solution-2}
\int_{0}^{\infty} dt  \int_{-\infty}^{\infty} dr
\left( y(r,t) \frac{\partial \phi(r,t)}{\partial t}  + \frac12 y(r,t)^2
 \frac{\partial \phi(r,t)}{\partial r} \right)  +
  \int_{-\infty}^{\infty} dr \ y(r,0)  \phi(r,0)   = 0,
\end{eqnarray}
This equation has no explicit derivatives of $y(r,t)$. By definition, a weak solution
to PDE~(\ref{eq:Burgers}) is a function $y(r,t)$ which satisfies the integral equation~(\ref{eq:weak-solution-2})
for any smooth and
bounded test function $\phi(r,t)$. The weak solution $y(r,t)$ does not have to be
differentiable everywhere. Any strong solution is also a weak solution  but not vice versa.
A weak solution to the FRG equation can exist and describe infrared physical quantities even when
the flow equation does not have a strong solution~\cite{Aoki:2014}. The same arguments are also applied to the
PME~\cite{Vazquez:2007}.

\hypertarget{sec:E}{}
\section{E. The nonanalytic fixed point: phase plane formalism} \label{sec:phase-plane}

We now look for the BSSs of the form~(6) to the PME~(5). The corresponding profile functions $F(\zeta)$
satisfy the nonlinear ODE
\begin{eqnarray}
\label{eq:for-F-1}
F(\zeta)\left[ F''(\zeta) +\frac1{\zeta} F'(\zeta)\right]
+ F'(\zeta)^2
-\delta \zeta F'(\zeta)  +(2\delta-1) F(\zeta) = 0.
\end{eqnarray}
In order to solve the ODE~(\ref{eq:for-F-1}) we will use the phase-plane formalism~\cite{Vazquez:2007}.
We introduce the phase variables $Z$~and~$Y$,
\begin{eqnarray}
\label{eq:phase-plane}
&&\!\!\!\!\!\!\!\!\!\!\!\! F(\zeta)=-\zeta^2 Z(\zeta), \ \ \  \ Y(\zeta)=-(2Z(\zeta)+\zeta Z'(\zeta)). \ \ \
\end{eqnarray}
The phase variables satisfy the autonomous first order differential equations
\begin{eqnarray}
\label{eq:for-Z}
&&  \dot{Z} +2 Z + Y =0, \\
&&  Z \dot{Y}  -  \dot{Z}(\delta-Y)+(4Y-1)Z=0, \label{eq:for-V}
\end{eqnarray}
where $\dot{X} \equiv d X/ d\ln \zeta $. Dividing Eq.~(\ref{eq:for-V})  by Eq.~(\ref{eq:for-Z})
we obtain an equation for $Y(Z)$
\begin{eqnarray}
\label{eq:Phase-plane-Diez-5-n1}
&&   \frac{d Y}{d Z} = \frac{  Z (2\delta-1)+
 2 Y Z+ (\delta-Y)Y }{ Z[2  Z+  Y]}.
\end{eqnarray}
Note, however, that $Y(Z)$ is not a single value function~\cite{Gratton:1990} for arbitrary $\delta$.
The relation between $\zeta$ and $Z$ is then given by
\begin{eqnarray}
\label{eq:Phase-plane-Diez-5-n12}
&& \frac{d \ln \zeta }{d Z} =- \frac{1}{2 Z + Y}.
\end{eqnarray}
Any solution of Eq.~(\ref{eq:Phase-plane-Diez-5-n1}) determines an integral curve in the
phase plane  $(Z,Y)$, which represents a BSS of a certain kind.
Thus, any FP of the FRG equation~(4) corresponds to a particular integral
curve in phase plane~$(Z,Y)$. A single integral curve passes through any regular point
of the phase plane while a curve corresponding to a FP of the FRG flow has to satisfy
certain boundary conditions at its ends.
To find out  which  curve corresponds to a physical FP
one needs to know the behavior in the vicinity of singular points
of system~(\ref{eq:for-Z})-(\ref{eq:for-V}).
\begin{figure}
\includegraphics[width=7.5cm]{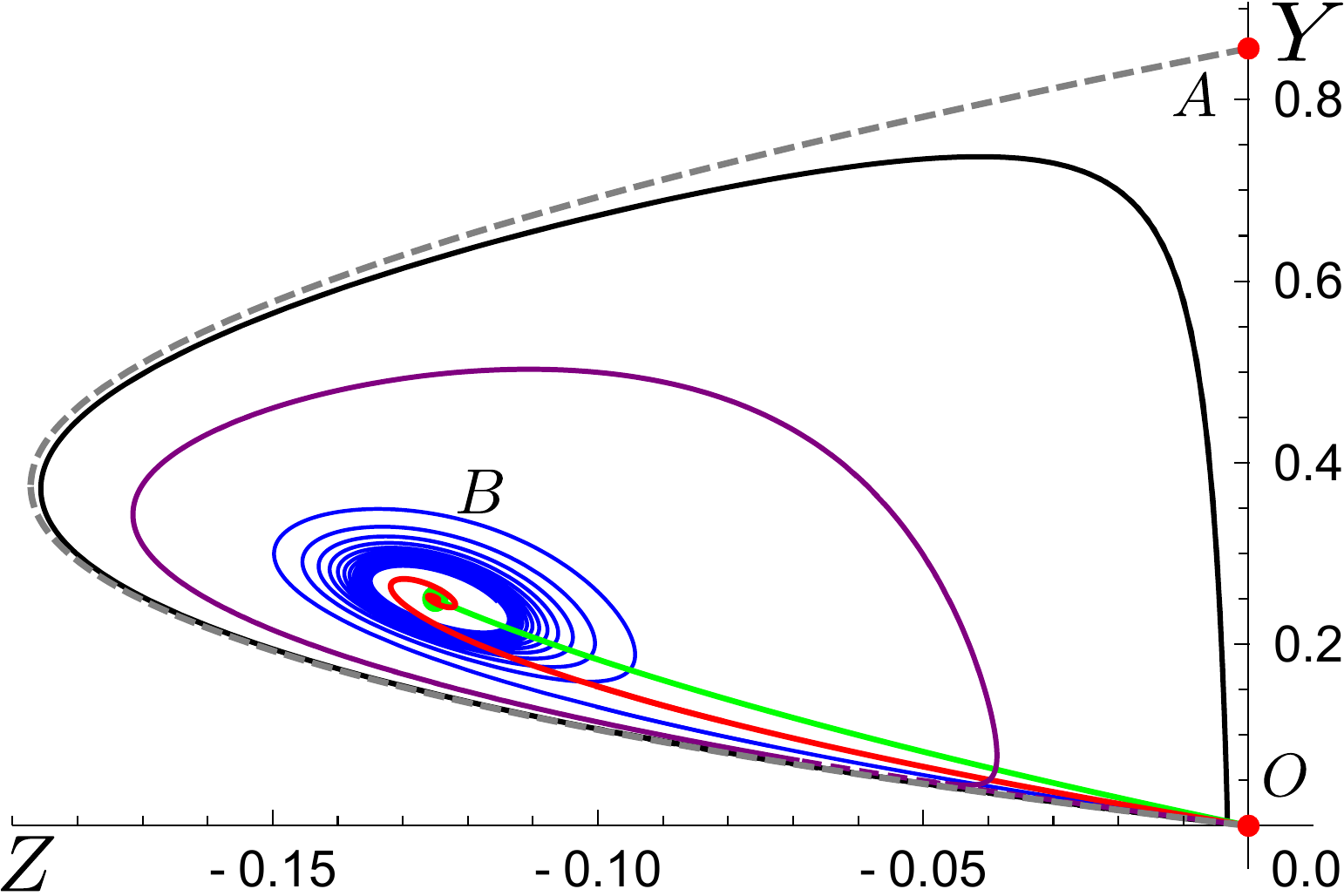} \ \ \ \ \ \ \ \ \ \
\includegraphics[width=8.3cm]{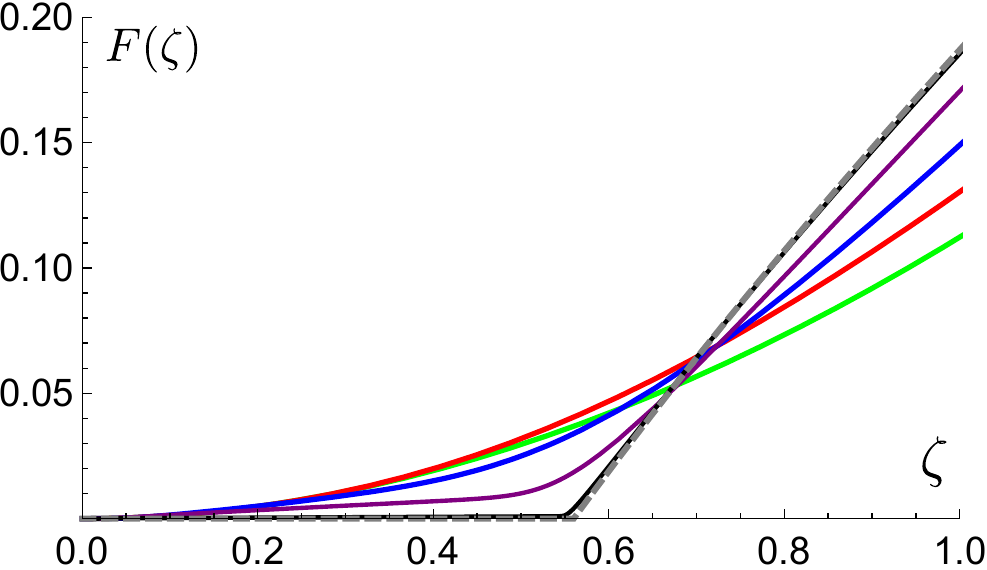}
 \caption{(\textbf{Left panel}) Curves in the phase plane $(Z,Y)$ describing the nonanalytic FP
 for different values of $\delta$:
  $\delta=\delta_c$ - dashed grey; $\delta=0.86$ - black;
  $\delta=0.9$ - purple  ; $\delta=1$ - blue ; $\delta=1.2$ - red;   $\delta=2$  green.
  (\textbf{Right panel}) Functions $F(\zeta)$ corresponding to the integral curves shown in
  the left panel and using the same color scheme.}
 \label{fig:phase-plane-SM}
\end{figure}
Since $F(\zeta)>0$ we are interested in the region $Z\le 0$
which is shown in the left panel of Fig.~\ref{fig:phase-plane-SM}.
In this region the
system~(\ref{eq:for-Z})-(\ref{eq:for-V}) has three singular points $O$: $(0,0)$, $A$: $(0,\delta)$ and
$B$: $(-\frac18,\frac14)$. It is easy to see that the singular point $B$ is itself
an integral curve which corresponds to the canonical GN FP independently of $\delta(\varepsilon)$
and which is given by
\begin{eqnarray}
\label{eq:canonicla-FP}
F(\zeta) = \frac{\zeta^2}8, \ \ \ \ w(\theta)=  \frac{1}8 \varepsilon \theta^2 \ \ \ \ \ (\tilde{N}=0).
\end{eqnarray}
The singular points $A$ and $B$ control the behavior of $F(\zeta)$ at small $\zeta$,
while the point $O$ describes
its asymptotic  behavior for $\zeta \to  \infty$. It turns out that
the new nonanalytic FP corresponds to the integral curve which connects
singular point $B$ (or limiting cycle around it) and point $O$ (see Fig.~\ref{fig:phase-plane-SM}).
However, numerical integration of Eqs.~(\ref{eq:for-Z}) and (\ref{eq:for-V})
shows that this curve exists only for $\delta>\delta_c \approx 0.8563265$.
At $\delta=\delta_c$ the curve passes from $O$ directly to the third singular point $A$ instead of spiraling
endlessly around point $B$.
Linearizing the flow (\ref{eq:for-Z})-(\ref{eq:for-V})
around point $B$ we find that there are three critical values of $\delta$:
$\delta_{\pm}=1\pm1/\sqrt{2}$ and $\delta_0=1$.

(i) The point $B$ is an unstable node for $\delta>\delta_{+}$, so that the curve goes
straight to the point $B$. In this case $Y(Z)$ is a single-valued function which could be found from
numerical integration of Eq.~(\ref{eq:Phase-plane-Diez-5-n1}). In this case it has a Taylor expansion
around $B$ which is given by
\begin{eqnarray}
\label{eq:flow-psi-17w1}
&& B: \ \ \   Y(Z)= \frac{1}{4} - \alpha_1
   \left(Z+\frac{1}{8}\right)+ O\left(\left(Z+\frac{1}{8}\right)^2\right) \ \ \ \mathrm{with } \ \ \
   \alpha_1= - 2 \left(\sqrt{2} \sqrt{2 \delta ^2-4 \delta +1}-2 \delta +1\right).
\end{eqnarray}
Although $Y(Z)$ is a single-value analytic function for $\delta>\delta_{+}$ the corresponding
FP is nonanalytic. Substituting (\ref{eq:flow-psi-17w1}) into
Eq.~(\ref{eq:Phase-plane-Diez-5-n12}) and integrating the resulting ODE we find that the
resulting FP functions $Z(\zeta)$
and $F(\zeta)$ are nonanalytic at the origin
\begin{eqnarray}
\label{eq:flow-psi-17w2}
&& Z(\zeta) \approx -\frac{1}8 +C_1 \zeta^{-2+\alpha_1}, \ \ \ \ \
 F(\zeta) \approx  \frac{1}8 \zeta^2 - C_1 \zeta^{\alpha_1}
 \ \ \ \ \mathrm{for} \ \ \ \ \ \zeta \to 0.
\end{eqnarray}
Here $C_1$ is an arbitrary constant which comes from the
integration of  Eq.~(\ref{eq:Phase-plane-Diez-5-n12}). While the function $Y(Z)$ is unique for any fixed
$\delta\ge\delta_c$, the function $F(\zeta)$ has a free parameter which fixes the value
of $C_1$ in Eq.~(\ref{eq:flow-psi-17w2}). This is related to existence of zero eigenvalue in the stability spectrum
that is discussed in the next section.

(ii) The point $B$ is an unstable focus  for  $\delta_0<\delta<\delta_{+}$, so that the integral curve
spirals finite time  towards $B$ until reaching it. In this case $Y(Z)$ is a multi-valued function so that
the functions $Z(\zeta)$ and $F(\zeta)$ are strongly nonanalytic at the origin.

(iii) The point $B$ is a stable focus for $\delta_{c}<\delta<\delta_0$. The integral curve
spirals infinitely many times around $B$ reaching a limiting cycle. This insures a very strong
nonanalyticity of the functions $Z(\zeta)$ and $F(\zeta)$  at the origin.

For $\delta_c<\delta$ the point $B$ represents
the solutions which asymptotically satisfy boundary condition $F(\zeta) \sim \zeta^2$ at $\zeta \to 0$,
but the function  $F(\zeta)$ is not analytic at $\zeta=0$. In order
to impose non-vanishing  boundary conditions for $F(\zeta)$
at large  $\zeta$ the integral curve has to pass through the point $O$ which
represents the behavior for $\zeta \to \infty$. For $Z<0$ it is a saddle point so that it can be reached
only along a single direction that determines the  asymptotic  behavior
of $F(\zeta)$ at infinity.
Expanding around $O$ we obtain
\begin{eqnarray}
\label{eq:Phase-plane-Diez2-6}
&&  O: \ \ \  Y(Z) = \frac{(1-2
   \delta ) Z}{\delta  } + \frac{2(2 \delta -1)^2 Z^2
   }{\delta ^3 }+ O(Z^3), \label{eq:Phase-plane-Diez-7}  \\
&& \mbox{}\hspace{15mm} Z(\zeta) \approx C_2 \zeta^{-1/\delta},\ \ \ F(\zeta) \approx C_2 \zeta^{2-1/\delta}
\ \ \ \ \mathrm{for} \ \ \ \ \ \zeta \to \infty, \label{eq:Phase-plane-Diez-7W}
\end{eqnarray}
where $C_2$ could be related to $C_1$ by matching of both asymptotic
solutions (\ref{eq:flow-psi-17w2}) and (\ref{eq:Phase-plane-Diez-7W}).
The integral curves connecting the point $O$ with point $B$ (or a limit cycle around $B$) are
shown in  the left panel of Fig.~\ref{fig:phase-plane-SM} for several values of
$\delta \ge \delta_c$ corresponding to different regimes. The corresponding
function $F(\zeta)$ representing the new nonanalytic FP of the FRG is shown in the right panel of
Fig.~\ref{fig:phase-plane-SM}. The function $F(\zeta)$
is  strongly nonanalytic at the origin
but surprisingly  grows  monotonically despite the fact that $Z(\zeta)$ could be oscillating function (see
Fig.~1 in the main text).
Once $\delta$ goes to $\delta_c$ the integral curve approaches the curve connecting
the singular points $O$ and $A$. It corresponds to the self-focusing solution for the initial function $u$
with a finite hole in the support around the origin which shrinks to zero in a finite time $T$.
There is no such nonanalytic FP for $\delta<\delta_c$, i.e. above  the
critical dimension $d_{\mathrm{c}} \approx 3.4$ (notice that it is only a one-loop estimation).

\hypertarget{sec:F}{}
\section{F. Stability of the nonanalytic fixed point} \label{sec:stability}

We now study the linear stability of the BSS solution $F(\zeta)$ using
the flow equation~(7) which we recall here
\begin{eqnarray}
\label{eq:pressure-ansatz1Q-2}
\partial_{\tau}F(\zeta;\tau) =
F(\zeta;\tau)\left[ F''(\zeta;\tau) + \frac1{\zeta} F'(\zeta;\tau)\right] + F'(\zeta;\tau)^2 -\delta \zeta F'(\zeta;\tau)
+(2\delta-1) F(\zeta;\tau),
\end{eqnarray}
where  we define a new time $\tau=-\ln(T-t)$ which has nothing to do with the imaginary time in the
action~(1).
To that end we linearize Eq.~(\ref{eq:pressure-ansatz1Q-2}) around the BSS $F(\zeta)$ by adding perturbation $\phi$
as $F(\zeta;\tau) =  F(\zeta)+ \phi(\zeta) e^{\lambda\tau}$
and expanding to the linear order in $\phi$. This gives
\begin{eqnarray}
\label{eq:flow-psi-1}
F(\zeta) \phi''(\zeta) +
\left[\frac{F(\zeta)}{\zeta} +2 F'(\zeta) - \delta \zeta \right] \phi'(\zeta)
+ \left[2\delta-1 -\lambda +  F''(\zeta) + \frac{F'(\zeta)}{\zeta} \right] \phi(\zeta) = 0.
\end{eqnarray}
By inspecting the different terms in Eq.~(\ref{eq:flow-psi-1}) we find that the eigenfunctions should
satisfies boundary conditions $\phi(0)=0$ and $\phi(\zeta) \sim \zeta^{(2\delta-1-\lambda)/\delta}$ for
$\zeta \to \infty$. However, it turns out that without additional conditions
the spectrum of Eq.~(\ref{eq:flow-psi-1}) is continuous
except for the case $\delta=\delta_c$. The focusing solution with $\delta=\delta_c$ (see the grey dashed
line in the right panel of Fig.~\ref{fig:phase-plane-SM}) belongs to
the so-called BSSs of the second kind  in
the classification of Barenblatt and Zeldovich~\cite{Vazquez:2007}.  In this case the boundary
conditions  for $\phi(\zeta)$  deduced directly from the flow equation~(\ref{eq:flow-psi-1}) are enough
to render the spectrum discrete.
The stability of this BSS of the second kind has been already studied in the mathematical
literature~\cite{Aronson:2003,Angenent:2001,Betelu:2000}.
Despite the fact that this solution is a marginal case in our problem let us outline the known results
in order to get intuition about stability in the general case. It was shown that this BSS
describes the generic disappearance of holes in the support (for 2D case)
even if one starts from a  non-radial
initial configuration $u(\vec{r},t=0)$. Notice that here
we need a weaker stability since the radial symmetry is imposed by the problem. It was found that
the spectrum has only three non-negative eigenvalues. The eigenvalue $\lambda=1$ is related to the property
that shifting the focusing time $T$ one arrives at another BSS.
The eigenvalue $\lambda=\delta_c$ corresponds to a
non-radial perturbation (forbidden in our problem) which shifts the focusing point from the origin
$\vec{r}=0$ to a finite $\vec{r}$. The eigenvalue $\lambda=0$ corresponds to the fact that one
can redefine $\zeta$ as $\zeta=r/[b(T-t)^{\delta_c}]$ with arbitrary $b$ and there is a freedom in the
choice of $b$.

For a BSS of first kind ($\delta>\delta_c$)
the boundary condition at $\zeta=0$ has to be imposed using additional physical arguments~\cite{Wiese:2016}.
The natural choice is considering the perturbations which resemble the presence of higher order
cumulants in the bare disorder distribution.
Since $F(\zeta) = \varepsilon  w' \left(\zeta^2/(2\varepsilon^2) \right)$
we impose
\begin{eqnarray}
\label{eq:perturbations}
\phi_n(\zeta) = \zeta^n f_n(\zeta), \ \ \ \ n = 2,4,6,...
\end{eqnarray}
where functions $f_n(\zeta)$ are bounded for $\zeta \to 0$ but not necessarily analytic.
Conditions~(\ref{eq:perturbations}) render the spectrum discrete.

Before we consider the whole spectrum let us first show that some eigenfunctions $\phi(\zeta)$ and
the corresponding eigenvalues $\lambda$ can be found from general
symmetry considerations similar to the case of BSS of second kind.  Indeed, $u(r,t)$
given by Eq.~(6) is a BSS solution to the PME~(5) for arbitrary
waiting time $T$. Thus, we have that
\begin{eqnarray}
\label{eq:flow-psi-2}
u(r,t) \to  u(r,t) + \frac{\partial u(r,t)}{\partial T} \Delta T
\end{eqnarray}
is also a BSS for small $\Delta T$.
Computing the derivative in Eq.~(\ref{eq:flow-psi-2}) we find
\begin{eqnarray}
\label{eq:flow-psi-3}
 \frac{\partial u(r,t)}{\partial T}  &=& (T-t)^{2\delta-2} \left[ (2\delta-1) F(\zeta;\tau)
 -\delta \zeta \partial_{\zeta}  F(\zeta;\tau)   - \partial_{\tau}  F(\zeta;\tau) \right] \nonumber  \\
&& = (T-t)^{2\delta-1} \left[ (2\delta-1) F(\zeta;\tau)
 -\delta \zeta \partial_{\zeta}  F(\zeta;\tau)   - \partial_{\tau}  F(\zeta;\tau) \right] e^{\tau}.
\end{eqnarray}
Substituting the BSS  $F(\zeta;\tau) \to F(\zeta)$ we identify the eigenvalue $\lambda_2=1$ which
corresponds to the eigenfunction
\begin{eqnarray}
\label{eq:flow-psi-4}
\phi_2(\zeta) = (2\delta-1) F(\zeta)  -\delta \zeta  F'(\zeta).
\end{eqnarray}
It is easy to check that the function~(\ref{eq:flow-psi-4}) indeed has the asymptotic behavior
$\phi_2(\zeta) = \zeta^2 f_2(\zeta)$ where $f_2(\zeta)$ is bounded but nonanalytical at $\zeta=0$

The FRG equation~(4) has a property that if $w(\theta)$ is a FP then
$b^2 w(\theta/b)$ is also a FP. Thus, the FRG equation has a line of FPs parameterized by $b$.
In general this could result in scaling behavior with non-universal values of critical exponents. However,
as we will see below the values of the critical exponents do not change along the line and they are universal
at least to one-loop order. In terms of the PME this symmetry implies that
if $u(\zeta;\tau)$ is a BSS to the PME  then $b^2 u( {r}/{b}; t)$ is also a BSS.  Using the
infinitesimal form of this transformation
\begin{eqnarray}
\label{eq:flow-psi-2bb}
u(r b^{-1},t) \to u(rb^{-1},t) + \frac{\partial u(r b^{-1},t)}{\partial b} \Delta b
\end{eqnarray}
we find
\begin{eqnarray}
\label{eq:flow-psi-5}
 \frac{\partial u(r b^{-1},t)}{\partial b}  &=& (T-t)^{2\delta-1} \frac{\partial  }{\partial b}
 b^2F\left(\zeta b^{-1}; \tau\right)  = b (T-t)^{2\delta-1}  \left[2 F \left( \zeta b^{-1}; \tau \right)
 - F'\left( \zeta b^{-1};\tau\right) \zeta b^{-1}  \right].
\end{eqnarray}
Thus, the flow has zero eigenvalue $\lambda=0$  which corresponds to the  eigenfunction
\begin{eqnarray}
\label{eq:flow-psi-6}
\phi (\zeta) = 2 F(\zeta)  -  \zeta  F'(\zeta).
\end{eqnarray}
Note, however, that the eigenfuction~(\ref{eq:flow-psi-6}) does
not fulfil the additional condition~(\ref{eq:perturbations}) and thus does not affect the stability properties.

In order to study the properties of the whole spectrum we now rewrite Eq.~(\ref{eq:flow-psi-1})
in terms of phase-variable $Z,Y$ as
\begin{eqnarray}
\label{eq:flow-psi-7}
-\zeta^2 Z(\zeta) \phi''(\zeta) +
\left[-\zeta Z(\zeta)  +2 \zeta Y(\zeta) - \delta \zeta \right] \phi'(\zeta)
+ \left[2\delta-1 -\lambda +  2Y(\zeta)+ \zeta Y'(\zeta) \right] \phi(\zeta) = 0,
\end{eqnarray}
where $\phi' = \partial_\zeta \phi(\zeta)$. It is convenient to introduce to $l=\ln\zeta$
and   $\dot{\phi} = \partial_l \phi(l)$ that gives Eq.~(9), i.e.
\begin{eqnarray}
\label{eq:flow-psi-8}
&&- Z \ddot{\phi} + \left[2 Y  - \delta  \right] \dot{\phi}
+ \left[2\delta-1 -\lambda +  2Y+  \dot{Y} \right] \phi = 0.
\end{eqnarray}

\textit{(i) Instability of the canonical FP}. -- We now reproduce the stability spectrum
of the canonical FP computed in the main text using scaling dimensions of the composite operators $\theta^{\tilde{n}}$
by solving Eq.~(\ref{eq:flow-psi-8}) with conditions~(\ref{eq:perturbations}). Substituting
$\phi(\zeta) \sim \zeta^{2\tilde{n}-2} $ with  $\tilde{n}=2,3,4,...$  into Eq.~(\ref{eq:flow-psi-8}) and using
that $Z_B=-\frac18$, $Y_B=\frac14$  and $\delta=(1+\varepsilon)/(2\varepsilon)$ we obtain
\begin{eqnarray}
\label{eq:stab-can-1}
&&\frac18 (2\tilde{n}-2)^2 + \left[\frac12  - \delta  \right] (2\tilde{n}-2)
+ \left[2\delta-1 -\lambda_{\tilde{n}} +  \frac12 \right]  = 0.
\end{eqnarray}
Solving Eq.~(\ref{eq:stab-can-1}) we find that
$\varepsilon \lambda_{\tilde{n}}= 2+\varepsilon - \tilde{n}(1+\varepsilon) + \frac{\varepsilon \tilde{n}^{ 2}}2  $
which coincides with the scaling dimensions
$[w^{(\tilde{n})}(0)]$ found in the main text in the limit of $N \to 0$. Thus
the canonical FP is indeed infinitely unstable in $d>2$.

\textit{(ii) Stability of the nonanalytic FP}. --  We now check the stability of the nonanalytic FP.
To that end we  switch from $\phi(l)$, $Z(l)$ and $Y(l)$ to $\phi(Z)$ and $Y(Z)$. Using that
\begin{eqnarray}
&& \dot{\phi} = \frac{\partial \phi}{\partial l} = \dot{Z} \frac{\partial \phi}{\partial Z}, \ \ \ \ \ \
 \ddot{\phi} = \frac{\partial^2 \phi}{\partial l^2}
= \ddot{Z} \frac{\partial \phi}{\partial Z} + (\dot{Z})^2\frac{\partial^2 \phi}{\partial Z^2},
\end{eqnarray}
%%%%%%%%%%%%%%%%%%%%%%%%%%%%%%%%%%%%%%%%%
\begin{figure}
\includegraphics[width=7cm]{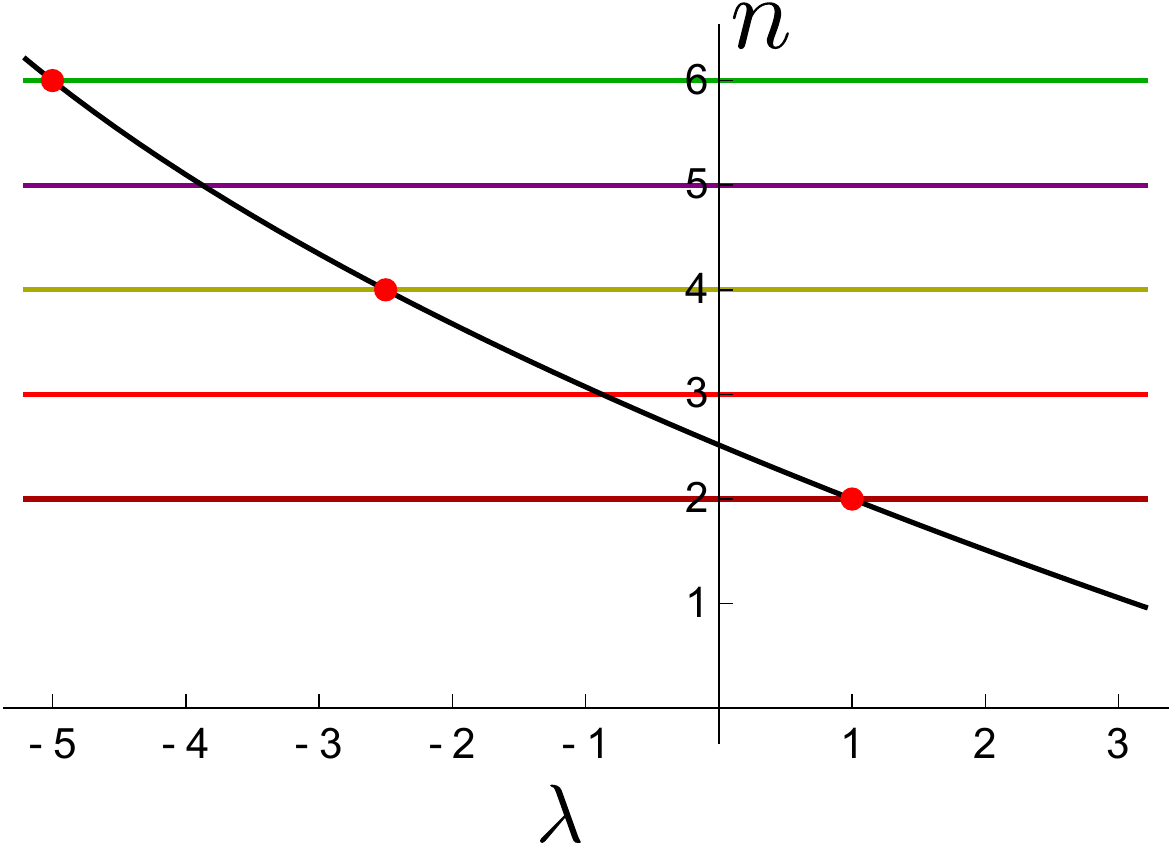} \ \ \ \ \ \ \ \ \ \
\includegraphics[width=7.5cm]{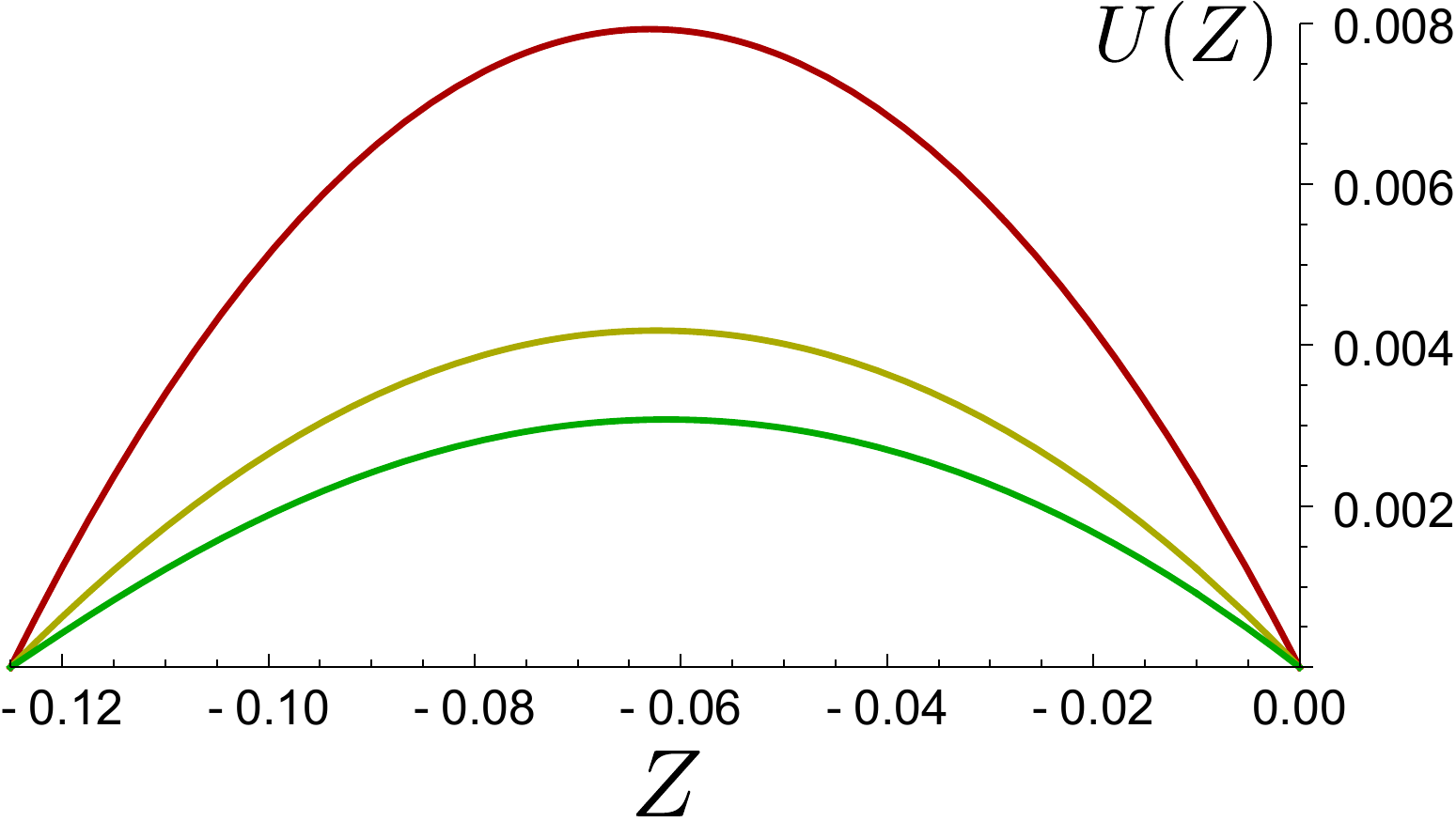}
 \caption{(\textbf{Left panel}) The stability eigenvalues computed using
  the condition~(\ref{eq:flow-psi-20}) for $\delta=3$.
  (\textbf{Right panel}) The corresponding eigenvectors $U(Z)$ using the same color scheme.}
 \label{fig:U-sev-lambda}
\end{figure}
%%%%%%%%%%%%%%%%%%%%%%%%%%%%%%%%%%%%%%%%%%%%%%%
we can rewrite Eqs.~(\ref{eq:flow-psi-8}) as
\begin{eqnarray}
\label{eq:flow-psi-10}
&&- Z ( 2 Z + Y)^2 \frac{\partial^2 \phi(Z)}{\partial Z^2}
- \left[ 4 Z^2 + 8 Y Z    -  Z   + Y^2
  \right]\frac{\partial \phi(Z)}{\partial Z}
+ \left[    \frac{Y}{Z} (Y - \delta) -\lambda  \right] \phi(Z) = 0,
\end{eqnarray}
which should be solved simultaneously with Eq.~(\ref{eq:Phase-plane-Diez-5-n1})
We now introduce the (inverse) logarithmic derivative
of $\phi(Z)$ as $U(Z) = \frac{\phi(Z)}{\phi'(Z)}$. Then we arrive at the systems of three first order
ODEs
\begin{eqnarray}
\label{eq:flow-psi-14}
&&   \frac{d Y(Z) }{d Z} = \frac{  Z (2\delta-1)+ 2 Y(Z) Z + (\delta-Y(Z))Y(Z) }{ Z[2  Z+  Y(Z)]}, \\
&&\frac{d U(Z) }{d Z} =  1+ \frac{4 Z^2 + 8 Y(Z) Z    -  Z   + Y(Z)^2  }{Z ( 2 Z + Y(Z))^2} U(Z) -
\frac{ Y(Z) (Y(Z) - \delta) -\lambda Z }{Z^2 ( 2 Z + Y(Z))^2}U(Z)^2,  \\
&& \frac{d \ln \zeta }{d Z} =- \frac{1}{2 Z + Y(Z)}.
\end{eqnarray}
Note that functions $Y(Z)$ and $U(Z)$ are single valued only for $\delta>\delta_{+}$ so that the stability
analysis is more simple in this case. For the sake of simplicity we restrict our consideration here
to this case, but the obtain conclusions are also applied for $\delta_c<\delta<\delta_{+}$.
We first study the asymptotic behavior of the eigenfunctions at small and large $\zeta$.
Expanding around  $O$, i.e. for  $\zeta \to \infty$ and thus $Z,Y \to 0$, we find
\begin{eqnarray}
\label{eq:flow-psi-16}
&& O :  U(Z) = \frac{Z}{ \lambda +1-2 \delta}-\frac{Z^2 \left(16 \delta ^3-8 \delta ^2 (\lambda +1)-4
   \delta +\lambda ^2+2 \lambda +2\right)}{\delta ^2 ( \lambda +1-2 \delta)^2} + O(Z^3), \\
&&\mbox{} \ \ \  \ \ \phi(Z) \approx C_3  Z^{-(2\delta-1-\lambda)}, \ \ \
\phi(\zeta) \approx C_3'  \zeta^{(2\delta-1-\lambda)/\delta}
\end{eqnarray}

Expansion around point B ($Z_B = -\frac{1}{8}$ $V_B = \frac14$),
i.e. for  $\zeta \to 0$,
which exists only for $\delta>\delta_{+}$, reads
\begin{eqnarray}
\label{eq:flow-psi-18}
&& B: \ \ \  U(Z) =  \beta_1 \left(Z+\frac{1}{8}\right) + O\left(\left(Z+\frac{1}{8}\right)^2\right), \\
&& \ \ \ \  \ \  \beta_1 =\frac{1}{1 + \delta  \left(2 \delta-3\right) +( \delta-\frac12)\sqrt{4 (\delta -2)
 \delta +2}
 - \frac{\sqrt{4 (\delta -2) \delta +2\lambda +2}}{
   \sqrt{4 (\delta -2) \delta +2}+2-2 \delta}}, \\
&& \ \ \ \  \ \  \phi(Z) \approx C_4 \left( Z+\frac18\right)^{1/\beta_1}, \ \ \
   \phi(\zeta) \approx C'_4 \zeta^{(\alpha_1-2)/\beta_1}, \label{eq:flow-psi-19}
\end{eqnarray}
where $C_4$and $C_4'$ are related by a cumbersome formula.
The condition~(\ref{eq:perturbations}) can be now written as
\begin{eqnarray}
\frac{\alpha_1-2}{\beta_1}=n, \ \ \ n=2,4,6... \label{eq:flow-psi-20}
\end{eqnarray}
One can check that
$(\alpha_1-2)/\beta_1=2$ for  $\lambda=1$ and any $\delta>\delta_{+}$, that is in consistence with
Eq.~(\ref{eq:flow-psi-4}).
The first several eigenfunctions  $U(Z)$ and eigenvalues computed numerically
for $\delta=3$ ($\varepsilon=0.2$)
using Eqs.~(\ref{eq:flow-psi-14}) are shown in Fig.~\ref{fig:U-sev-lambda}.

\begin{figure}
\includegraphics[width=7.5cm]{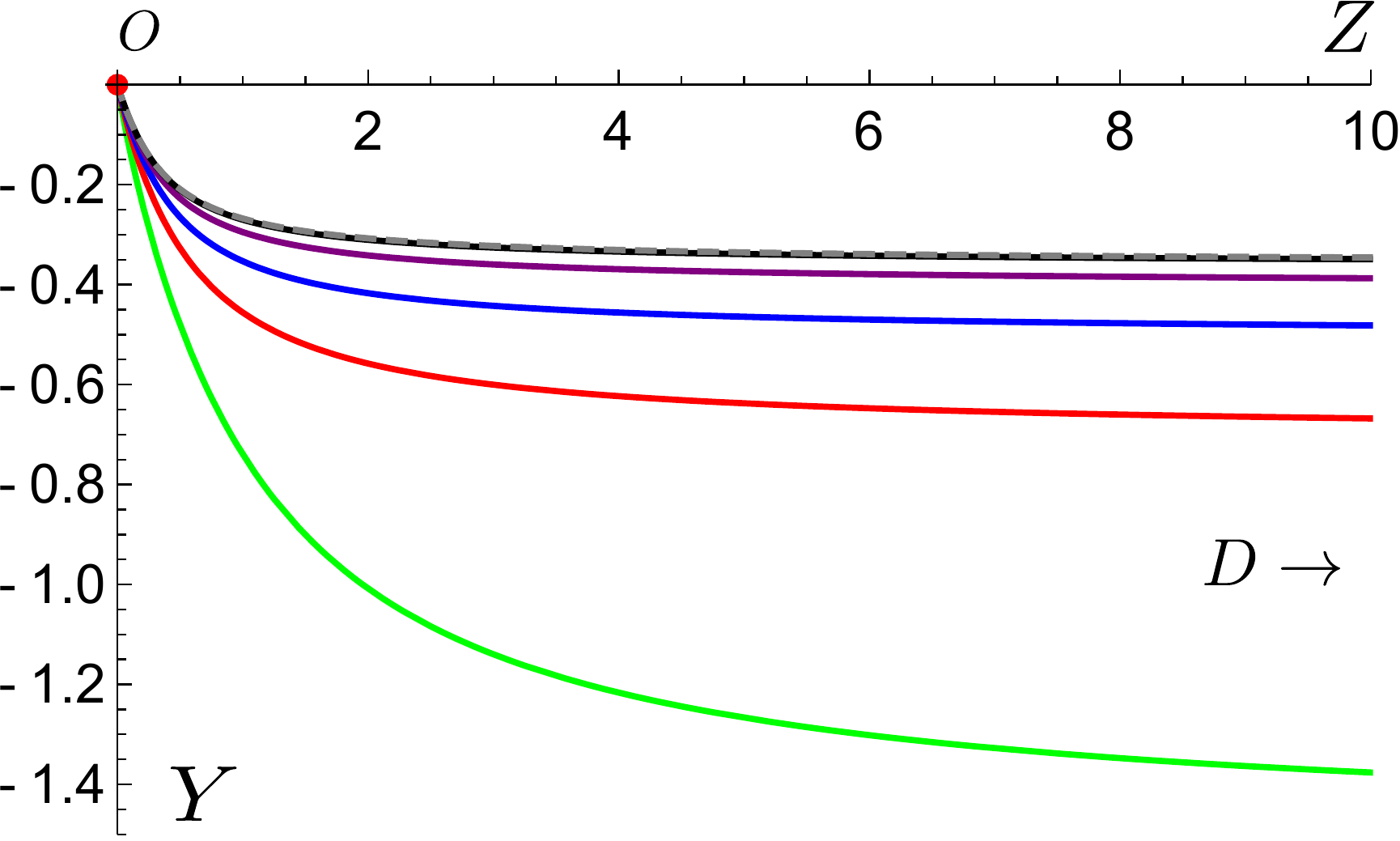} \ \ \ \ \ \ \ \ \ \
\includegraphics[width=9cm]{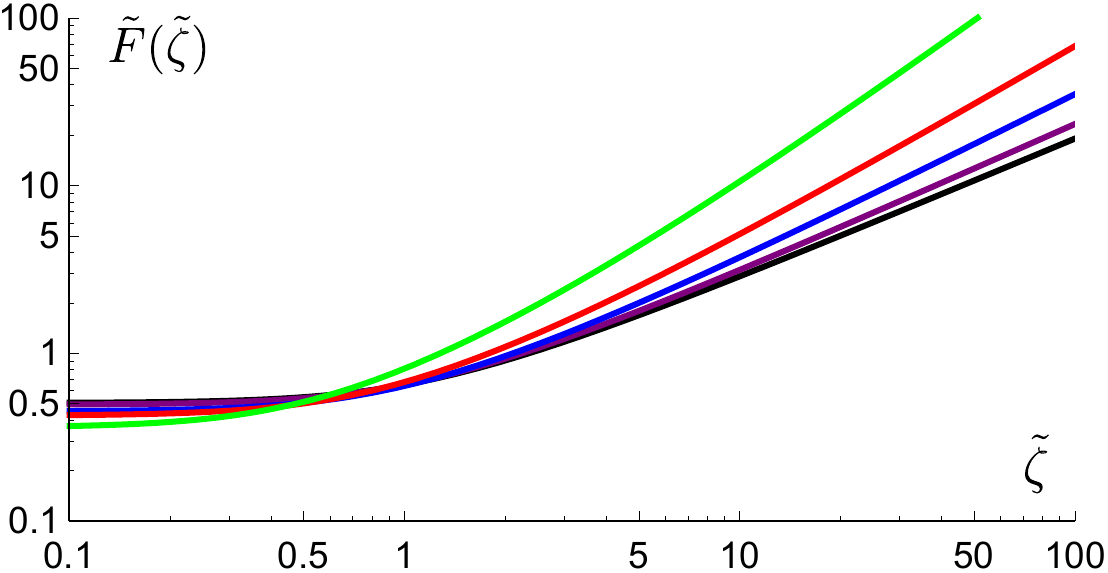}
 \caption{(\textbf{Left panel}) Curves in the phase plane $(Z,Y)$ describing the postfocusing
 FSS  for different values of $\delta$:
  $\delta=\delta_c$ - dashed grey; $\delta=0.86$ - black;
  $\delta=0.9$ - purple  ; $\delta=1$ - blue ; $\delta=1.2$ - red;   $\delta=2$  green.
  (\textbf{Right panel}) Functions $\tilde{F}(\tilde{\zeta})$ corresponding to the integral curves shown in
  the left panel (using the same color scheme).}
 \label{fig:post-focus}
\end{figure}

\hypertarget{sec:G}{}
\section{G. The postfocusing regime} \label{sec:post-focus}

In the postfocusing regime, i.e. for $T<t<T_0$, the time evolution of the system
can be described by a FSS of type~(11) such that $\tilde{F}(\tilde{\zeta}=0)>0$. Thus,
this regime describes the diffusive phase of the relativistic fermions. The
corresponding profile function $\tilde{F}(\tilde{\zeta})$ is the solution to
\begin{eqnarray}
\label{eq:for-FSS-F-1}
\tilde{F}(\tilde{\zeta})\left[ \tilde{F}''(\tilde{\zeta}) +\frac1{\tilde{\zeta}} \tilde{F}'(\tilde{\zeta})\right]
+ \tilde{F}'(\tilde{\zeta})^2
+\delta \tilde{\zeta} \tilde{F}'(\tilde{\zeta})  -(2\delta-1) \tilde{F}(\tilde{\zeta}) = 0.
\end{eqnarray}
Equation~(\ref{eq:for-FSS-F-1}) can be analyzed using the same phase variables $Z$ and $Y$ but now
in the semiplane $Z>0$. The phase variables satisfy the same systems
of autonomous first order differential equations~(\ref{eq:for-Z})
and (\ref{eq:for-V}), however, the relation with $\tilde{F}(\tilde{\zeta})$  is now given by
\begin{eqnarray}
\label{eq:phase-plane-FSS}
&&\!\!\!\!\!\!\!\!\!\!\!\! \tilde{F}(\tilde{\zeta})
=\tilde{\zeta}^2 Z(\tilde{\zeta}), \ \ \  \
Y(\tilde{\zeta})=-(2Z(\tilde{\zeta})+\tilde{\zeta} Z'(\tilde{\zeta})).
\end{eqnarray}
The FSS in the postfocusing regime corresponds to the integral curve connecting
the singular point $O$,  which describes the asymptotics for $\tilde{\zeta} \to \infty$,
and the singular point $D$, ($Z_D=\infty$, $Y_D = (1-2\delta)/2$), which describes
the behavior at $\tilde{\zeta}=0$. Expansion around point $O$ is still given
by Eq.~(\ref{eq:Phase-plane-Diez-7}) with $Z>0$ while expansion around point $D$ reads
\begin{eqnarray}
\label{eq:expansion-around-D-1}
D: \ \ \ \ Y(Z)=\frac{1}{2} (1-2 \delta )+\frac{8 \delta ^2-6 \delta +1}{16 Z}
-\frac{(2 \delta -1) (1-4 \delta )^2}{192 Z^2}
+\frac{\delta  (2 \delta -1) (1-4 \delta )^2}{1536 Z^3} + O(Z^{-4}).
\end{eqnarray}
Using these expansions we numerically integrate Eqs.~(\ref{eq:for-Z}) and (\ref{eq:for-V}).
The resulting integrals curves in the plane $(Z,Y)$ and the corresponding profile functions
$\tilde{F}(\tilde{\zeta})$ in the postfocusing regime are shown in Fig.~\ref{fig:post-focus}
for several values of $\delta$. We now can compute the explicit form of FSS to the PME which describes
the  FRG flow in the diffusive metal phase
\begin{eqnarray}
&& F(\tilde{\zeta}) = 1+ \frac{1}{4} (2
   \delta -1) \tilde{\zeta} ^2     -\frac{1}{64} (2 \delta -1) (4 \delta -1) \tilde{\zeta} ^4
    +\frac{1}{576} (2 \delta -1) (4 \delta -1) (5   \delta -2) \tilde{\zeta} ^6 \nonumber  \\
&&   -\frac{(2 \delta -1) (4 \delta -1) \left(144 \delta ^2-116 \delta +23\right) \tilde{\zeta}
   ^8}{24576}
   +\frac{(2 \delta -1) (4 \delta -1) \left(4024 \delta ^3-4882 \delta ^2+1952
   \delta -257\right) \tilde{\zeta }^{10}}{921600} + O(\tilde{\zeta}^{12}). \nonumber  \\
\end{eqnarray}
We can now derive the characteristic function of the local DOS distribution~\cite{Horovitz:2002}
\begin{eqnarray}
W(\Theta) = (T_0-T)^{2\beta+1} \int_0^{\sqrt{2\Theta}/(T_0-T)^\delta} \tilde{\zeta} F(\tilde{\zeta})
d\tilde{\zeta}.
\end{eqnarray}
We find that in the diffusive metal phase the $n$th cumulant of the DOS distribution scales as
$\overline{\rho^n(0)}^{c}\sim(T_0-T)^{\beta-2\delta(n-1)} $. Thus, developing a singularity in the FRG flow
at $T=T_0$ provides a unique mechanism for the generating of a finite averaged DOS in the
diffusive metal phase.

\end{document}